\documentclass[11pt]{article}
\usepackage{cite}
\usepackage{amsmath,amssymb,amsfonts}
\usepackage{breq n}
\usepackage{algorithmic}
\usepackage{graphicx}
\usepackage{textcomp}
\usepackage{xcolor}
\def\BibTeX{{\rm B\kern-.05em{\sc i\kern-.025em b}\kern-.08em
    T\kern-.1667em\lower.7ex\hbox{E}\kern-.125emX}}

\usepackage{geometry}
\newlength{\logowidth}
\newlength{\logohspace}
\begin{document}

	\title{Asymptotic regime analysis of NOMA uplink networks under QoS delay Constraints}
	\date{September 2019}
	\maketitle
	\vspace{1cm}
		\begin{center}
\author{MSc Project Report BELLO Mouktar}
	
	\vspace{0.5cm}	
	
	\normalsize{Master 2 Systemes Intelligents Communicants option Signal et Telecommunications, \\year 2018-2019}\\
		
	\vspace{0.5cm}	
	
	\normalsize{ETIS - ENSEA / Universit\'e de Cergy-Pontoise / CNRS UMR 8051}\\
	\normalsize{6 avenue du Ponceau, 95014 Cergy-Pontoise Cedex, France}
	
	\vspace{1cm}
	
	 Supervisor \\
	CHORTI Arsenia, Associate Professor at the ENSEA \\
	\vspace{0.5cm}
    Monitoring officer \\ LUZZI Laura, Assistant Professor, ENSEA
	\vspace{0.5cm}

\end{center}

\newpage
\tableofcontents
\newpage

\begin{abstract}
In the fifth generation and beyond (B5G) technologies, delay constrains emerge as a topic of particular interest for ultra reliable low latency communications (e.g., enhanced reality, haptic communications). In this report, we study the performance of a two-user uplink non-orthogonal multiple access (NOMA) network under quality of service (QoS) delay constraints, captured through each user's delay exponents in their effective capacity (EC). We propose novel closed-form expressions for the EC of the NOMA users and validate them through Monte Carlo simulations. Interestingly, our study shows that in the high signal to noise ratio (SNR) region, the ``strong'' NOMA user has a limited EC no matter how large the transmit SNR is, under the same delay constraint as the ``weak'' user. We show that for the weak user OMA achieves higher EC than NOMA at small values of the transmit SNR $\rho$ and that NOMA become more beneficial at high values of $\rho$. For the strong user, we show that NOMA achieves a higher EC than OMA at small values of the transmit SNR and that at high values of $\rho$ OMA becomes more beneficial. By introducing user pairing when more than two NOMA users are present, we show that NOMA with user-pairing outperforms OMA in term of the total link layer EC. Finally, we find the set of pairs which gives the highest total link-layer in the uplink for NOMA with multiple user-pairs.
\end{abstract}

\newpage
\textbf{ACRONYMS}\\
\textbf{ETIS}  Information Processing and Systems Teams \\
\textbf{NOMA}  Non Orthogonal Multiple Access\\
\textbf{OMA}  Orthogonal Multiple Access\\
\textbf{EC}  Effective Capacity\\
\textbf{TDMA} Time Division Multiple Access \\
\textbf{SC} Superposition Coding \\
\textbf{SIC Successive Interference Cancellation}  \\
\textbf{AWGN}  Additive White Gaussian Noise\\

\textbf{Notation}\\
\textbf{$B(a,b,)$}  Beta function\\
\textbf{$\Gamma(a)$}  Gamma function\\
\textbf{$\mathbb{E}[.]$}  Expected value\\
\textbf{$\log_2(.)$}  Binary logarithm\\
\textbf{$U(a,b,z)$}  Confluent hypergeometric function of the second kind\\
\textbf{$W_{u,v}(z)$} Whittaker function\\

\newpage
I did my MSc project (stage de M2 en recherche) at the ETIS laboratory, from April to September 2019.
\section{Presentation of ETIS research laboratory \cite{ensea}}
ETIS, Information Processing and Systems Teams, is a joint research unit at CNRS (UMR 8051), ENSEA Cergy and the University of Cergy-Pontoise. ETIS is attached mainly to the Institute of Computer Science and their Interactions (INS2I). ETIS is located on the premises of ENSEA and UCP (St-Martin 1). The laboratory hosts faculty members, researchers and administrative and technical staff from these three institutions.
Researchers in the laboratory report to CNRS section 7 "Information Science: Signals, Images, Languages, Automation, Robotics, Interactions, Integrated Hardware-Software Systems".
\newline
ETIS has over a hundred members and is structured into four research teams:
\begin{itemize}
  \item Multimedia Indexing and Data Integration (MIDI)\newline
  The work of the MIDI team focuses on indexing, searching and searching large amounts of data, ranging from relational databases, to heterogeneous Web data (XML, RDF, information flow) and multimedia data (images, videos, 3D objects).\newline
The activity of the team is organized around two axes: masses of data and
multimedia research systems
  \item Information, Communications, Imaging (ICI) \newline
  The research activities of the ICI team revolve around four axes: error correcting coding, information theory, resource allocation and
     imagery.
  \item CELL, Formerly Architectures, Systems, Technologies for Embedded Reconfigurable Units (ASTRE)\newline
  The theme of the Cell team focuses on heterogeneous systems-on-a-chip reconfigurability methods, with the originality to explore their adaptability at different levels, circuit, system and software, for many embedded applications (video, telecommunication, radiology, embedded systems for health).
  \item Neurocybernetic \newline
  The goal of the neuro team is to model the neural mechanisms and important brain structures of infant development to make autonomous robots that can perceive the outside world, learn for themselves and interact with it.
\end{itemize}

\section{MSc Project: Background Concepts}
Non orthogonal multiple access (NOMA) schemes have attracted a lot of attention recently, allowing multiple  users to be served simultaneously with enhanced spectral efficiency; it is known that the boundary of achievable rate pairs using NOMA is outside the capacity region achievable with orthogonal multiple access techniques \cite{islam2016power} or other approaches to increase spectral efficiency \cite{Chorti_FTN, Chortisemidefinite, Chortioverview}. Superior achievable rates are attainable though the use of superposition coding at the transmitter and of successive interference cancellation  (SIC) at the receiver \cite{yu2018link}, \cite{saito2013non}, \cite{Rihem}, while it is shown in \cite{Wenjuan_secrecy} that NOMA is also advantageous with respect to physical layer security \cite{Chorti_secFTN, Chorti_Hollanti, Chorti_helping, Chorti10}. The SIC receiver decodes multi-user signals with descending received signal power and subtracts the decoded signal(s) from the received multi-user signal, so as to improve the signal-to-interference ratio. The process is repeated until the signal of interest is decoded. In uplink NOMA networks, the strongest user's signal is decoded first (as opposed to downlink NOMA networks in which the inverse order is applied).

Besides, in a number of emerging applications, quality of service (QoS) delay constraints become increasingly important, e.g., ultra reliable low latency (URLLC) systems.  Furthermore, in future wireless networks, users are expected to necessitate flexible delay guarantees for achieving different service requirements. Henceforth, in order to satisfy diverse delay requirements, a simple and flexible delay QoS model is imperative to be applied and investigated. In this respect, the effective capacity (EC) theory can be employed \cite{yu2016tradeoff},\cite{wu2003effective} \cite{tang2007cross}, \cite{tian2018performance}, with EC denoting the maximum constant arrival rate which can be served by a given service process, while guaranteeing the required statistical delay provisioning.

Applying the EC theory in a NOMA setting with multiple users, the $m$-th user's EC over a block-fading channel, is defined as:
\begin{align}
       E_c^m = -\frac{1}{\theta_m T_{\text{f}} B} \ln \left( \mathbb{E}\left[e^{-\theta_m T_{\text{f}} B R_m}\right] \right) \quad \left(\text{in b/s/Hz}\right),\label{eq:EC2}
\end{align}
where $T_f$ is the symbol period, $B$ the block bandwidth and $\mathbb{E}\left[\cdot\right]$ denotes expectation over the channel gains.

The rest of the report is organized as follows: In Section 3 we present multiple access techniques, in Section 4 the theory of effective capacity (EC), in Section 5 the application of TEC in the case of two-user network uplink OMA and NOMA, in Section 6 we introduce user-pairing and multiple NOMA pairs and in Section 7 we present our simulation results.

\section{Multiple Access Techniques}
Multiple Access techniques are classified into two major families: orthogonal and non-orthogonal.
\subsection{Orthogonal Multiple Access (OMA)}
In OMA, each user can exploit orthogonal communication resources within either a specific time slot (time division multiple access – TDMA), a frequency band (frequency division multiple access – FDMA), or a code (code division multiple access – CDMA), to avoid multiple access interference.

 \begin{figure}[t]
 \begin{center}
      \includegraphics[width=0.5\textwidth]{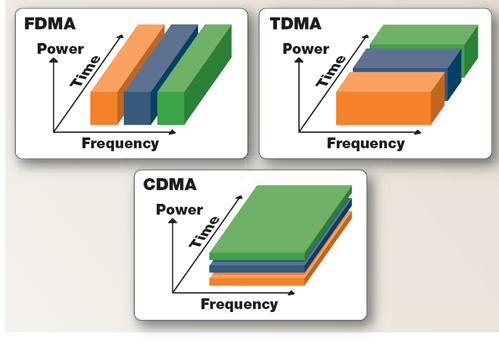}
      \caption{Fundamental OMA techniques \cite{WinNT}}
      \label{fig}
      \end{center}
  \end{figure}
 Limitations of OMA:
 \begin{itemize}
     \item  Low spectral efficiency and throughput of the system (user is given a resource irrespective of his channel conditions)
     \item Limited in term of number of users.
 \end{itemize}
 In order to overcome these limitations, the NOMA technique has been proposed.
\subsection{Non Orthogonal Multiple Access}
In the literature, there are two categories of NOMA: the power domain NOMA which attains multiplexing in power domain, and the code domain NOMA which attains multiplexing in code domain \cite{islam2016power}. \cite{bloch2013strong}. See the Figure 2. In this report we are going to focus only on power domain NOMA.
 \begin{figure}[ht]
 \begin{center}
      \includegraphics[width=10cm]{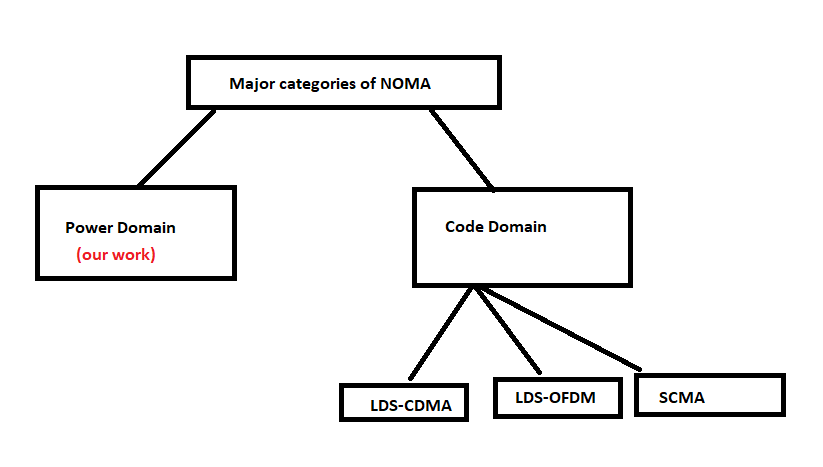}
      \caption{Different types of NOMA}
      \label{fig}
      \end{center}
  \end{figure}
     \subsubsection{Power Domain NOMA}
Unlike orthogonal access techniques, in NOMA users do not share the available resource. The entire resource is used by all users. Multiplexing is done in the power domain, by giving each user a specific power coefficient.
This class of NOMA relies primarily on two techniques:  superposition coding (SC) and  successive interference cancellation (SIC).
 \begin{itemize}
     \item Superposition coding  (SC)\newline
The superposition coding is a technique of simultaneously communicating information to several receivers by a single source.\cite{islam2016power} This operation is done in two stages:
First, bits representing the message are mapped into a sequence of bits using a specific point-to-point encoder for each user. In the case of two users, we have $S_1$ and $S_2$, when using encoders: $E_1$:$\lbrace0,1\rbrace^{\lfloor2 T R_1 \rfloor}$$\rightarrow C_{1}^T$ and $E_2$:$\lbrace0,1\rbrace^{\lfloor2 T R_2 \rfloor}$$\rightarrow C_{2}^T$. $R_1$ and $R_2$ represent the transmission rates of user 1 and user 2. $T$ is the block length at the output of the encoders. $C_{i}$ is the code library of the user i. Then, a summation device provides an output sequence to broadcast to all users by the source, i.e., a base station (BS) in a downlink scenario:

\begin{equation}
    X=\sqrt{P \alpha_{1}}S_{1}+\sqrt{P \alpha_{2}}S_{2},
\end{equation}
where $\alpha_i$ is a fraction of the total power allocated to user $i$, subject to the
 constraint $\alpha_{1}+\alpha_{2}=1$, in the case of a two users network ( in general, $\sum_{i=1}^{M}\alpha_{i}=1$, with $M$ the total number of users) \cite{islam2016power}. And $P$ is the total power. Then, the respective powers of user 1 and user 2 are $P_1$=$\alpha_1 P$ and $P_2$=$\alpha_2 P$.

     \item Successive interference cancellation\newline
The idea of SIC is for the receiver to retrieve a message from the received signal from a superposition of other users' messages. After a user's signal is decoded, it is subtracted from the combined signal before decoding the next user's signal in the decoding SNR order. When SIC is applied, one user signal is decoded considering other users' signals as interference. The operation decode-subtract is repeated by the receiver until the intended message is decoded.
\newline
 \begin{figure}[ht]
 \begin{center}
      \includegraphics[width=15cm]{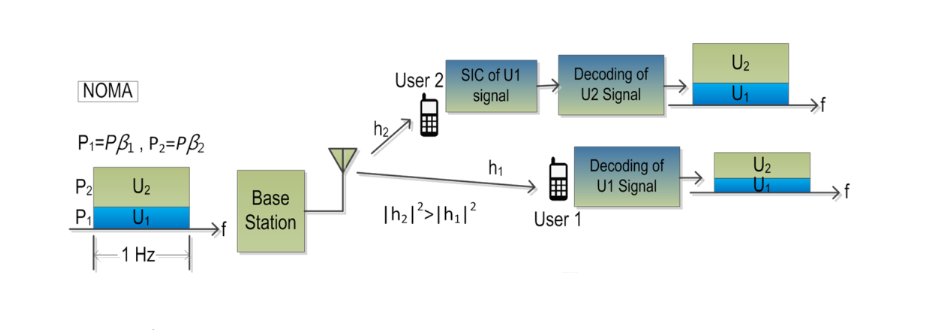}
      \caption{Downlink NOMA \cite{islam2016power}}
      \label{fig}
      \end{center}
  \end{figure}
\newline
Assume a $M$ users NOMA network with users U$_1$, U$_2$,...,U$_N$ in a Rayleigh block fading channel with respective gains $|h_1|^2<|h_2|^2<...<|h_M|^2$, transmitting symbols $S_1, S_2,...,S_M$ respectively, with power
\(E[|S_i|^2]=P_i\), $i=1,2,...,M$. 

In the downlink scenario, the SIC operation is done by the users' terminals. And in the case of two users, the process is as following:\newline
1) At user 1, the weak user, using $D_{1}:C_{1}^{T}\rightarrow\lbrace0,1\rbrace^{\lfloor2 T R_1 \rfloor}$, the message $S_1$ is decoded, considering the $\sqrt{P \alpha_{2}}S_{2}$ as a noise. \newline
2) At the user 2, the strong one, the messages of all weaker users are decoded successively and subtracted before the user retrieves it's message. From the received combined signal $Y_{2}=Z \times h_{2}$, user 2:\newline
a) Decode user 1 message $S_1$ by using the decoder, $D_{1}:C_{1}^{T}\rightarrow\lbrace0,1\rbrace^{\lfloor2 T R_1 \rfloor}$.\newline
b) Subtract this message from the combined signal it receives:\newline
 $\Tilde{Y_{2}}=Y_{2}-\sqrt{P \alpha_{1}}S_{1} h_{2}$ \newline
c) And then, using the decoder $D_{2}:C_{2}^{T}\rightarrow\lbrace0,1\rbrace^{\lfloor2 T R_2 \rfloor}$ it decodes it's own message.
\newline

 \begin{figure}[ht]
 \begin{center}
      \includegraphics[width=15cm]{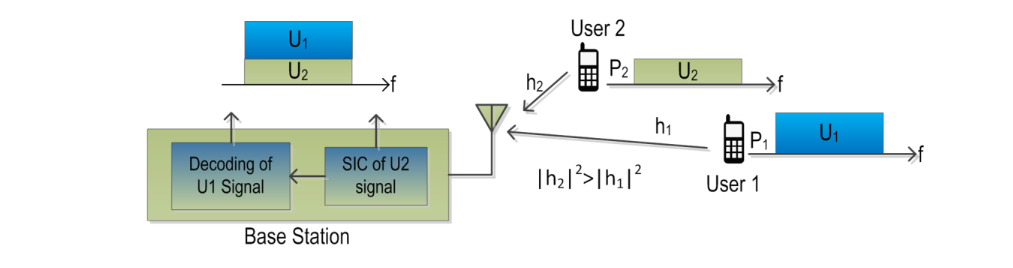}
      \caption{Uplink NOMA \cite{islam2016power}}
      \label{fig}
      \end{center}
  \end{figure}

In the uplink scenario, which interests me in this report, the SIC is performed at the base station (BS). For a two user NOMA uplink network, the received signal at the BS is: \newline
\begin{equation}
    Z= h_{1}\sqrt{P \alpha_{1}}S_{1}+h_{2}\sqrt{P \alpha_{2}}S_{2} + W
\end{equation}
As mentioned above, $P \alpha_{i}$ represents the transmission power of user $i, i=1,..,M$. And $W$ is an additive white Gaussian noise (AWGN) random variable with zero mean and variance $\sigma^2$. Here the BS decodes and cancels successively the signals from the strong to weak user, i.e., in descending order of the channel coefficients.

 \end{itemize}

In the rest of this report, power domain uplink NOMA will be refered to as NOMA for simplicity.
\section{Theory of Effective Capacity}

In the uplink, the BS will first decode the symbol of the strongest user treating the transmission of the weaker one as interference. After decoding it suppresses it from $Z$ and will decode the signal of the next strongest user. Following the SIC principle, and denoting by \(\rho = \frac{1}{\sigma^2}\) the transmit SNR, the achievable rates, in b/s/Hz, for the $m^{th}$ user is expressed as \cite{fan2015uplink}:

 \begin{equation}
    R_m= \log_2 \left(1+ \frac{\rho P_m|h_m|^2}{1+\rho \sum_{l=1}^{m-1}P_l|h_l|^2}\right).
    \label{Rm}
 \end{equation}

Replacing \eqref{Rm} into \eqref{Ecdef}, we obtain the following expression for the EC of the $m^{th}$ user:
\begin{eqnarray}
E_c^m &=& \frac{1}{\beta_m}\log_2\left(\mathbb{E}\left[(1+ \frac{\rho P_m|h_m|^2}{1+\rho \sum_{l=1}^{m-1}P_l|h_l|^2})^{\beta_m}\right]\right),
\end{eqnarray}
where $\beta_m=-\frac{\theta_mT_fB}{\ln{2}}$ is the normalized (negative) QoS exponent for $m=\{1,2,..,M\}$.\newline

To account for the ordering of the channel gains we make use of the theory of order statistics in the following analysis \cite{yang2011order}.
Assume a network of $M$ users with $|h_1|^2<|h_2|^2<...<|h_M|^2$,  $x_m=|h_m|^2, m=1,2,...M$ to denote independent and identically distributed (i.i.d.) random variables with underlying probability density function (pdf) $f(x_{m})$. Ranking these random variables using  order statistics we obtain:
\begin{equation}
    f_{\gamma_{m:M}}(x) = \psi_m f(x) (1-F(x))^{M-m}F(x)^{m-1},
    \label{pdforder}
\end{equation}
where $f_{\gamma_{m:M}}$ denotes the pdf of the $m$-th ordered random variable in a population of $M$ (in the two-user case $M=2$),   \(\psi_m=\frac{1}{B(m,M-m+1)}\), and, $B(a,b)$ is the beta function $B(a,b)$=$\frac{\Gamma(a)\Gamma(b)}{\Gamma(a+b)})$, with  $\Gamma(a)=a!$. Finally, $F(x)$ denotes the cumulative density function (cdf) of the the random variable  $x_{m}$, $m=\{1,2\}$.

Assuming a Rayleigh wireless environment, the NOMA gains of the channel, denoted by $x_m$, are exponentially distributed with respective pdf and cdf respectively \cite{papoulis2002probability}
\begin{equation}f(x)=e^{-x},\text{   }
F(x)=1-e^{-x}.
\label{pdfcdf}
\end{equation}

\section{General Expression of the Effective Capacity}
The general expression of the EC of the $m^{th}$ user in a $M$ users uplink NOMA scenario can be expressed as:
\begin{equation}
E_c^m= \frac{1}{\beta_m}\log_2\Bigg(\int_{0}^{\infty} \int_{x_1}^{\infty} \int_{x_2}^{\infty}...\int_{x_{m-1}}^{\infty}(1+\frac{\rho P_{m}x_m}{1+\sum_{l=1}^{m-1}\rho P_{l}x_l})^{\beta_m} f_{\gamma_{1:2:3:...M}}(x_1,x_2,...,x_m)d_{x_m}\ d_{x_m-1}...d_{x_1}\Bigg),
\end{equation}
where $f_{\gamma_{1:2:3:...M}}(x_1,x_2,...,x_m)$ is the joint distribution of $x_i$, $i=1,2,..,m$.
\subsection{Case of a Two Users OMA}
We note that $P_1+P_2=P=1$ and omit the power coefficient in the OMA case. For the case of $M$ users using  OMA, e.g., the TDMA technique where each user transmits for $\frac{1}{M}$ of the time, we have that achievable data rate for the user $m$:
\begin{equation}
  \widetilde{R}_{m}=\frac{1}{M}\log_2\Big(1+\rho |h_m|^2\Big), m=1,2,..,M
\end{equation}

Then for the case of $M$ users OMA, the expression of the $m^{th}$ user's EC is given by:
\begin{equation}
\widetilde{E}_c^{m}=\frac{1}{\beta_m}\log_2\Big(\mathbb{E}\left[(1+\rho |h_m|^2)^{\frac{\beta_m}{M}}\right] \Big),m=1,2,..,M.
\end{equation}

For $M=2$ users we have:
\begin{equation}
\widetilde{E}_c^{1}=\frac{1}{\beta_1}\log_2\Big(\mathbb{E}\left[(1+\rho |h_1|^2)^{\frac{\beta_1}{2}}\right] \Big)
\end{equation}

\begin{equation}
\widetilde{E}_c^{2}=\frac{1}{\beta_2}\log_2\left(\mathbb{E}\left[(1+\rho |h_2|^2)^{\frac{\beta_2}{2}}\right] \right)
\end{equation}
Closed-form expressions of both users OMA uplink are provided in Appendix A.
\subsection{Case of Two users NOMA}

for $m=1,2$, M=2, using \eqref{pdforder} and \eqref{pdfcdf} we have the pdfs of the ordered channel gains:
\begin{align}
   f_{\gamma_{1:2}}(x_{1}) = \frac{1}{B(1,2)} f(x_{1}) (1-F(x_{1})) = 2 e^{-2 x_{1}}.
   \label{singlepdf}
  \end{align}
In \cite{yang2011order}, the joint distribution of two order statistics is given by:
\begin{equation}
f_{\gamma_{l:M},\gamma_{k:M}}(x,y)=\frac{M!}{(l-1)!(k-l-1)!(M-k)!}(1-F(x))^{l-1}f(x)\times(F(x)-F(y))^{k-l-1}f(y)(F(y))^{M-k}.
\end{equation}
Applying this in the case of two users, we get the joint distribution between user $1$ and user $2$, $  (M=2, l=1, k=2)$
\begin{equation}
    f_{\gamma_{1:2}}(x_1, x_{2}) = 2 f(x_{1})f(x_{2}) = 2 e^{-x_{1}}e^{-x_{2}} .
   \label{pdf2joint}
\end{equation}
In this context, the EC of User $i$, denoted by $E_c^i$ is expressed as:

\begin{equation}
    E_c^1 = \frac{1}{\beta_1}\log_2\left(\mathbb{E}[(1+ \rho P_1 x_{1})^{\beta_1}]\right)=\frac{1}{\beta_1}\log_2\left(\int_{0}^{\infty}(1+ \rho P_1 x_1)^{\beta_1}f_{\gamma_{1:2}}(x_1) dx_1\right) ,
    \label{EC1}
\end{equation}

\begin{equation}
  E_c^2 = \frac{1}{\beta_2}\log_2\left( \mathbb{E}[(1+ \frac{\rho P_2 x_{2}}{1+ \rho P_1 x_{1}})^{\beta_2}]\right)=\frac{1}{\beta_2}\log_2\left(\int_{0}^{\infty}\int_{x_{1}}^{\infty}\Big(1+ \frac{\rho P_2 x_2}{1+\rho P_1 x_1}\Big)^{\beta_2}f_{\gamma_{1:2}}(x_1,x_2) dx_2 dx_1\right).
  \label{EC2}
\end{equation}
By replacing respectively \eqref{singlepdf} in \eqref{EC1}; \eqref{pdf2joint} in \eqref{EC2} we obtain respectively closed-form expressions of the first and second user:
\begin{equation}
E_c^1 = \frac{1}{\beta_1}\log_2\left(\frac{2}{P_1 \rho}\times U(1,2+\beta_1,\frac{2}{\rho P_1})\right)
\end{equation}

\begin{eqnarray}
  E_c^2&=&
  \frac{1}{\beta_2}\log_2\Big(2 P_2^{1-\beta_2}(\rho P_2)^{\beta_2} e^{\frac{1}{\rho P_2}}e^{-\frac{(P_1-P_2)}{\rho P_2}}\Big) +\frac{1}{\beta_2}\log_2\Big(\sum_{j=0}^{-\beta_2}\binom{-\beta_2}{j}(\rho P_1)^{j}\times \sum_{k=0}^{\infty}\frac{(-1)^k (P_2-P_1)^k}{k!}\frac{1}{1+j+k}\nonumber\\
  &\times&
  \Big[\Gamma[2+\beta_2+j+k,\frac{1}{\rho P_2}]-(\rho P_2)^{-1-j-k}\Gamma[1+\beta_2,\frac{1}{\rho P_2}] \Big] \Big)
  \label{eq4}
\end{eqnarray}
\newline
The proof is provided in Appendix B.
\newline
\subsubsection{Case 1 Delay-Constrained Users}

Lemma 1: Considering the individual EC in Uplink NOMA and OMA, for the both users we have that:
 a) When $\rho\rightarrow0$, $E_c^1\rightarrow0$, $E_c^2\rightarrow0$, $\widetilde{E}_c^{1}\rightarrow0$, $\widetilde{E}_c^{2}\rightarrow0$, $E_c^1-\widetilde{E}_c^{1}\rightarrow0$, $E_c^2-\widetilde{E}_c^{2}\rightarrow0$ \\
b) When $\rho\rightarrow +\infty$, $E_c^1\rightarrow+\infty$, $E_c^2\rightarrow \frac{1}{\beta_2}\log_2\left(\mathbb{E}\left[(1+\frac{P_2 |h_2|^2}{P_1 |h_1|^2})^{\beta_2}\right]\right)$, $\widetilde{E}_c^{1}\rightarrow+\infty$, $\widetilde{E}_c^{2}\rightarrow+\infty$, $E_c^1-\widetilde{E}_c^{1}\rightarrow+\infty$, $E_c^2-\widetilde{E}_c^{2}\rightarrow-\infty$

The proof is provided in Appendix C.

We notice that for both users, from  Lemma 1.(a) either in NOMA or OMA, their individual EC start at the same initial value of 0, at small values of $\rho$.
On the other hand, from  Lemma 1.(b), we notice that the EC of the stronger user is limited by $ \frac{1}{\beta_2}\log_2(\mathbb{E}[(1+\frac{P_2 |h_2|^2}{P_1 |h_1|^2})^{\beta_2}]) $ no matter how large the transmit SNR $\rho$ is. We believe that this is due to the fact that only  user 2 (stronger user) experiences interference. In fact, at the base station, the message of the user 2 is decoded by considering the signal of the user 1 (weak user) as interference. User 1, meanwhile, does not experience any interference, since it's message is decoded in the last position. On the contrary, for the weak user, when $\rho\rightarrow\infty$ it's achievable EC in NOMA uplink approaches infinity. This is the exact opposite of the downlink scenario, where it is the weaker user which is limited in terms of EC, when $\rho\rightarrow+\infty$. \cite{yu2018link} \cite{yu2015multi}

Now the question is how the effective capacity evolves over $\rho$ between these extreme values $\rho\rightarrow 0$ and $\rho\rightarrow +\infty$? To answer  this question and to further analyze the impact of $\rho$ on the individual EC, in a two-user uplink NOMA network and in a two-user OMA network, we investigate the derivative with respect to $\rho$ as in \cite{yu2018link}.\newline
Lemma 2: Considering the EC of user 1, in the uplink of a two-user NOMA network and OMA, we have that:
       a) At any values of $\rho$, $ \frac{\partial E_c^1}{\partial\rho}\ge0$, and $ \frac{\partial \widetilde{E}_c^{1}}{\partial\rho}\ge0$ \\
      b) When\footnote{The sign of this term depend on the sign of $(P_1-\frac{1}{2})$. And we work with $P_1=0.2$} $\rho\rightarrow0$, $\lim\limits_{\rho\rightarrow0}(\frac{\partial (E_c^1-\widetilde{E}_c^{1})}{\partial\rho})=\frac{(P_1-\frac{1}{2})}{\ln 2}\mathbb{E}[|h_1|^2]$ \\
   c) When $\rho$ is very large, $\frac{\partial (E_c^1-\widetilde{E}_c^{1})}{\partial\rho})\approx\frac{1}{2\rho \ln2}\ge0$ and it approaches 0 when $\rho\rightarrow\infty$.

       The proof is provided in Appendix C.

    Lemma 2.(b) indicates that, for the weaker user,user 1, when the transmit SNR $\rho$ is very small, the EC uplink OMA is increasing faster than in NOMA. On the other hand, Lemma 2.(c) shows that when the transmit SNR is very large, the EC uplink NOMA increases faster than in OMA. Combining Lemma 2 and Lemma 1, we can conclude that, $E_c^1-\widetilde{E}_c^{1}$ starts at the initial value of 0, first decreases, and at the end tends to $\infty$ with a gradually reducing slope. This means that for the weaker user, OMA achieves higher EC than NOMA at small values of the transmit SNR $\rho$. At high values of $\rho$, NOMA becomes more beneficial for the weaker user. Finally, when $\rho\rightarrow\infty$ the performance gain of NOMA over OMA becomes stable, for the weaker user.
    \newline
 Lemma 3: Considering the EC of user 2, in uplink two-user NOMA network and OMA, we prove that:
    a) At any values of $\rho$, $ \frac{\partial E_c^2}{\partial\rho}\ge0$, And $ \frac{\partial \widetilde{E}_c^{2}}{\partial\rho}\ge0$ \\
   b2) When\footnote{The sign of this term depend on the sign of $(P_2-\frac{1}{2})$. Here it's positive.} $\rho\rightarrow0$, $\lim\limits_{\rho\rightarrow0}(\frac{\partial (E_c^2-\widetilde{E}_c^{2})}{\partial\rho})=\frac{(P_2-\frac{1}{2})}{\ln 2}\mathbb{E}[|h_2|^2]$ \\
  c) When $\rho$ is very large, $\frac{\partial (E_c^2-\widetilde{E}_c^{2})}{\partial\rho}\approx-\frac{1}{2 \ln2}\frac{1}{\rho}<0$ and it approaches 0 when $\rho\rightarrow\infty$. \\

The proof is provided in Appendix D.

Lemma 3.(b) indicates that, for the stronger user,user 2, when the transmit SNR $\rho$ is very small, the EC uplink NOMA has a faster increasing slope than in OMA. On the other hand, Lemma 3.(c) shows that when the transmit SNR is very large the EC uplink OMA increases faster than in NOMA. Combining Lemma 3 and Lemma 1, we can conclude that, $E_c^2-\widetilde{E}_c^{2}$ starts at the initial value of 0, first increases, and at the end decreases to $-\infty$ with a gradually diminishing slope. This means that for the stronger user, NOMA achieves higher EC than OMA at small values of the transmit SNR $\rho$. At high values of $\rho$, OMA becomes more beneficial for the stronger user. Finally, when $\rho\rightarrow\infty$ the performance gain of OMA over NOMA becomes stable, for the stronger user.
    \newline
Lemma 4: Considering $V_N=E_c^1+E_c^2$, the total EC in a two users NOMA, We have that:
a) At any values of $\rho$, $ \frac{\partial V_N}{\partial\rho}\ge0$ \\
             b) When $\rho\rightarrow0$, $V_N\rightarrow0$, $ \lim\limits_{\rho\rightarrow 0}(\frac{\partial V_N}{\partial\rho})=\frac{P_1}{\ln2}\mathbb{E}[|h_1|^2]+\frac{P_2}{ \ln2}\mathbb{E}[|h_2|^2]\ge0$ \\
 c) When $\rho\rightarrow\infty$, $V_N\rightarrow\infty$, $\lim\limits_{\rho\rightarrow\infty}(\frac{\partial V_N}{\partial\rho})=0$.
   \\
    Considering $V_O=\widetilde{E}_c^{1}+\widetilde{E}_c^{2}$, the total EC in a two users system OMA, we have that: \\
             d) At any values of $\rho$, $ \frac{\partial V_O}{\partial\rho}\ge0$ \\
              e) When $\rho\rightarrow0$, $V_O\rightarrow0$, $ \lim\limits_{\rho\rightarrow 0}(\frac{\partial V_O}{\partial\rho})=\frac{1}{2 \ln2}\mathbb{E}[|h_1|^2]+\frac{2}{ \ln2}\mathbb{E}[|h_2|^2]\ge0$ \\
              f) When $\rho\rightarrow\infty$, $V_O\rightarrow\infty$, $\lim\limits_{\rho\rightarrow\infty}(\frac{\partial V_O}{\partial\rho})=0$.

The proof is provided in Appendix F.
\newline
Lemma 4.(b) indicates that when the NOMA scheme is applied, the total EC has a constant slope, at small values of the transmit SNR $\rho$; The exact value of the slope depends on the average of the channel power gains and the allocated power coefficients. Similarly from lemma 4.(e), we note that when using the OMA scheme, the total EC has a constant slope whose exact value depends only on the average of the channel power gains. On the other hand, when $\rho\rightarrow\infty$, lemma 4.(c) and lemma.(f) show that the increasing slope of the $V_N$ and $V_O$ gradually diminish.\newline

\subsubsection{Case 2 Delay-Unconstrained Users}

In this case, investigation of the EC of the two-user, uplink NOMA and OMA networks, is done without the delay constrain.  The impact of the transmit SNR $\rho$ in this case is also investigated.\newline

Lemma 5: Considering the EC for the weaker user with $\theta_1\rightarrow0$, in NOMA and OMA, we prove that:
a) When $\theta_1\rightarrow0$, $ \lim\limits_{\theta_1\rightarrow0}E_c^1=\mathbb{E}[R_1]$ ,
              $ \lim\limits_{\theta_1\rightarrow0}\widetilde{E}_c^{1}=\mathbb{E}[\widetilde{R}_1]$, $\lim\limits_{\theta_1\rightarrow0}(E_c^1-\widetilde{E}_c^{1})=\mathbb{E}[R_1]-\mathbb{E}[\widetilde{R}_{1}]$ \\
             b) When $\theta_1\rightarrow0, \rho\rightarrow\infty$ \\
              $\lim\limits_{\substack{\theta_1\rightarrow0\\\rho\rightarrow\infty}}E_c^1=\infty$, $\lim\limits_{\substack{\theta_1\rightarrow0\\\rho\rightarrow\infty}}\widetilde{E}_c^{1}=\infty$, $\lim\limits_{\substack{\theta_1\rightarrow0\\\rho\rightarrow\infty}}(E_c^1-\widetilde{E}_c^{1}) =\infty$.\\
Considering the EC for the stronger user with $\theta_2\rightarrow0$, in NOMA and OMA, we prove that:\\
             c) When $\theta_2\rightarrow0$, $ \lim\limits_{\theta_2\rightarrow0}E_c^2=\mathbb{E}[R_2]$\\
              $ \lim\limits_{\theta_2\rightarrow0}\widetilde{E}_c^{2}=\mathbb{E}[\widetilde{R}_2]$, $\lim\limits_{\theta_2\rightarrow0}(E_c^2-\widetilde{E}_c^{2})=\mathbb{E}[R_2]-\mathbb{E}[\widetilde{R}_{2}]$
              d) When $\theta_2\rightarrow0, \rho\rightarrow\infty$, \newline
              $\lim\limits_{\substack{\theta_2\rightarrow0\\\rho\rightarrow\infty}}E_c^2=\mathbb{E}[\log_2(1+ \frac{P_2 |h_2|^2}{P_1 |h_1|^2})]$, $\lim\limits_{\substack{\theta_2\rightarrow0\\\rho\rightarrow\infty}}\widetilde{E}_c^{2}=\infty$, $\lim\limits_{\substack{\theta_2\rightarrow0\\\rho\rightarrow\infty}}(E_c^2-\widetilde{E}_c^{2}) =-\infty$

The proof is provided in Appendix G.

From 5.(a) and 5.(c), we note that for both OMA and NOMA, when there is no delay requirement the individual achievable link-layer rates for both users are equal to their ergodic capacities. Same conclusions have in the past been drawn for the downlink, \cite{yu2018link}.
Furthermore, from Lemma 1 and from Lemma 5, we notice that we get a similar conclusions with or without delay requirements at high transmit SNRs.

In fact, from Lemma 1.(b) and Lemma 5.(b), we note that, with or without delay requirements, NOMA achieves higher EC than OMA at high transmit SNRs, for the weaker user. On the other hand, from Lemma 1.(b) and Lemma 5.(d) we note that the stronger user, in a NOMA uplink two-user system, can achieve only a limited EC at high transmit SNRs. But this limit is different according to whether there is a delay requirements or not. This means that, with or without delay requirements, at high SNRs, OMA achieves higher EC than NOMA for the stronger user.
 \section{Effective Capacity of Multiple NOMA Pairs}

  After analyzing the two-user NOMA network, we investigate the total achievable link-layer rate for multiple NOMA pairs in the uplink scenario. For $M$ users, we get $\frac{M}{2}$ groups with these following index $i=1,2,..,\frac{M}{2}$, where each group contains 2 users.
  Inside each group, NOMA is implemented, and for inter-group TDMA is assumed to be applied.

  The achievable data rate of the two users $1^{st}$ and $2^{nd}$ of the $i^{th}$group, where $|h_{1,i}|^2\leq|h_{2,i}|^2$ can be formulated as follows:
   \begin{equation}
    R_{1,i}= \frac{2}{M}\log_2 \left(1+ \rho P_{1,i}|h_{1,i}|^2\right)
     \label{R1i}
 \end{equation}
   \begin{equation}
    R_{2,i}=\frac{2}{M} \log_2 \left(1+ \frac{\rho P_{2,i}|h_{2,i}|^2}{1+\rho P_{1,i}|h_{1,i}|^2}\right).
    \label{R2i}
 \end{equation}

  On the other hand, if users 1 and 2 each have their message transmitted using TDMA, the achievable data rate is as follows (here we make use of the fact that $P=P_1+P_2=1$ and we omit the total power $P$):
    \begin{equation}
    R_{j}= \frac{1}{M}\log_2 \left(1+ \rho |h_{j}|^2\right), j\in \left\{1_i,2_i\right\}
    \label{Rj}
    \end{equation}
 $\frac{1}{M}$ means that each user has only one time slot to transmit. And $\frac{2}{M}$ is a resource available for the two users inside a NOMA group.

 By replacing \eqref{R1i} and \eqref{R2i} in (1) we get respectively the following ECs of the $1^{st}$ and $2^{th}$ user of the $i^{th}$ group:
 \begin{eqnarray}
    E_c^{1,i} &=& \frac{1}{\beta_{1,i}}\log_2\left(\mathbb{E}[(1+ \rho P_{1,i}|h_{1,i}|^2)^{\frac{2 \beta_{1,i}}{M}}]\right),
    \label{EC1i}
\end{eqnarray}

 \begin{eqnarray}
    E_c^{2,i} &=& \frac{1}{\beta_{2,i}}\log_2\left(\mathbb{E}[(1+ \frac{\rho P_{2,i}|h_{2,i}|^2}{1+ \rho P_{1,i}|h_{1,i}|^2})^{\frac{2 \beta_{2,i}}{M}}]\right),
    \label{EC2i}
\end{eqnarray}

On the other hand, replacing \eqref{Rj} in (1) we get the expressions for both users while using TDMA scheme:
 \begin{eqnarray}
   \widetilde{E}_c^{1,i} &=& \frac{1}{\beta_{1,i}}\log_2(\mathbb{E}[(1+ \rho |h_{1,i}|^2)^{\frac{\beta_{1,i}}{M}}]),
\end{eqnarray}

 \begin{eqnarray}
  \widetilde{E}_c^{2,i} &=& \frac{1}{\beta_{2,i}}\log_2(\mathbb{E}[(1+\rho |h_{2,i}|^2)^{\frac{\beta_{2,i}}{M}}]),
\end{eqnarray}
Here, we analyze the total EC of multiple NOMA pairs, denoted $W_N$, in comparison with the total EC for the $M$ OMA users, $W_O$,

\begin{equation}
W_N=\sum_{i=1}^{\frac{M}{2}}(E_c^{1,i}+E_c^{2,i}),
\end{equation}
and
\begin{equation}
    W_O=\sum_{i=1}^{\frac{M}{2}}(\widetilde{E}_c^{1,i}+\widetilde{E}_c^{2,i}).
\end{equation}

As in \cite{yu2018link}, to investigate the region of $\rho$ in which NOMA offers higher total link-layer rates EC compared to OMA, the following Lemma is provided.
\newline
Lemma 6: Considering the difference between the total EC of multiple NOMA pairs and the  OMA, we prove that:
a) When $\rho\rightarrow0$, $W_N-W_O$$\rightarrow0$, and \newline
              $\lim\limits_{\rho\rightarrow0} \frac{\partial(W_N-W_O)}{\partial\rho}=\sum_{i=1}^{\frac{M}{2}}\Big(\frac{2 P_{1,i}-1}{M \ln2}(\mathbb{E}[|h_{1,i}|^2]-\mathbb{E}[|h_{2,i}|^2)\Big)\geq0$\\
b) When $\rho\rightarrow\infty$, $W_N-W_O$$\rightarrow Q$, given in \eqref{Q}, and
             $\lim\limits_{\rho\rightarrow\infty} \frac{\partial(W_N-W_O)}{\partial\rho}=0$

where
\begin{equation}
 Q= \lim\limits_{\rho\rightarrow\infty}(W_N-W_O)=\sum_{i=1}^{\frac{M}{2}}\Big(\frac{1}{\beta_{1,i}}\log_2(P_{1,i}^{\frac{2 \beta_{1,i}}{M}} \mathbb{E}[(  |h_{1,i}|^2)^{\frac{\beta_{1,i}}{M}}])+\frac{1}{\beta_{2,i}}\log_2(\frac{\mathbb{E}[(1+ \frac{P_{2,i}|h_{2,i}|^2}{ P_{1,i}|h_{1,i}|^2})^{\frac{2 \beta_{2,i}}{M}}]}{\mathbb{E}[( |h_{2,i}|^2)^{\frac{\beta_{2,i}}{M}}]})\Big)
 \label{Q}.
\end{equation}

The proof is provided in Appendix H.

From Lemma 6, we can conclude that $W_N-W_O$ starts initially at 0, it fist increases at small values of transmit SNR $\rho$, and finally approaches a constant Q given in \eqref{Q}, at high values of the transmit SNR. That means multiple NOMA pairs outperforms OMA for low transmit SNRs.  And this performance gain of NOMA over OMA becomes stable when the transmit SNR is extremely high.

\section{Numerical Results}

In this section the proposed Lemmas in the previous sections will be validated through Monte Carlo simulations.
We considered a two user uplink NOMA system, with the following setting: transmission power of users, $P_1$=0.2, $P_2$=0.8 (i.e power coefficients $\alpha_1$=0.2, $\alpha_8$=0.8, and the total power $P$=1) and we set the normalized delay exponent equal to $\beta_1$=$\beta_2$=-1 for both users.

\begin{figure}[ht]
 \begin{center}
      \includegraphics[width=12 cm]{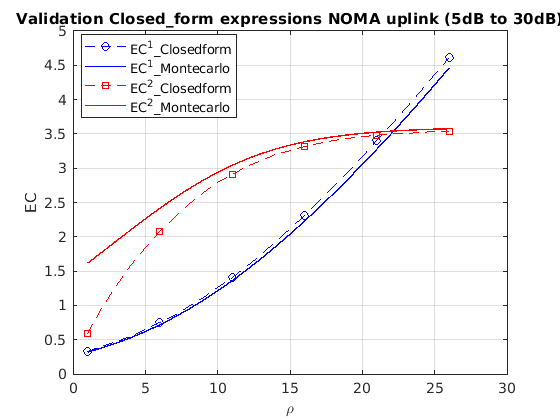}
      \caption{Validation of the closed-form expressions in uplink two-user NOMA system.}
      \label{fig5}
 \end{center}
  \end{figure}

To confirm the accuracy of the proposed closed-form expressions for EC in NOMA scheme for both users, we used Monte Carlo simulations. Figure \ref{fig5} shows the curves of the both users effective capacities, $EC^1$ and $EC^2$, using the closed-form expressions and using the Monte Carlo. So the accuracy of these closed-form can be confirmed, except in the very low SNR region (discrepancy).

\begin{figure}[ht]
  \begin{center}
      \includegraphics[width=12 cm]{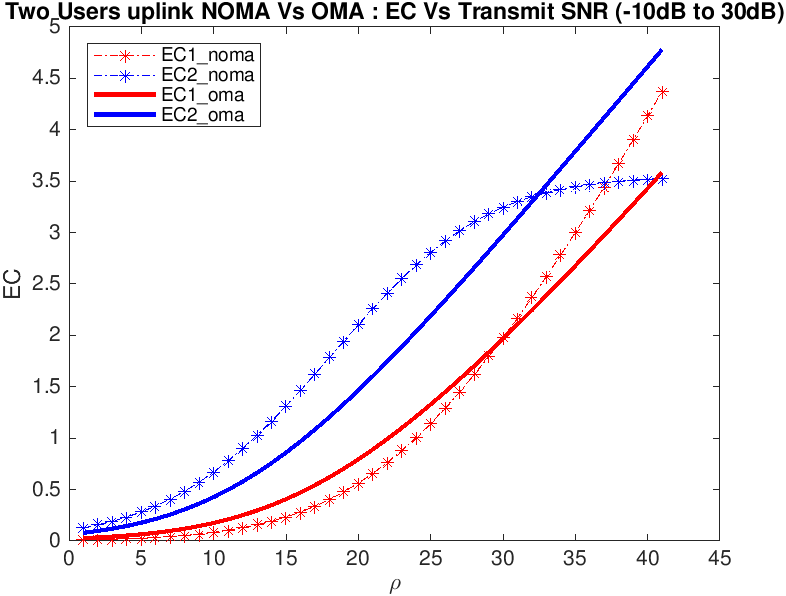}
      \caption{Effective capacities $E_c^1$ and $E_c^2$ in a two-user NOMA uplink network compared to Ecs of two users OMA, as functions of the transmit SNR $\rho$ (-10 to 30 dB).}
      \label{fig6}
    \end{center}
\end{figure}

\begin{figure}[h]
     \begin{center}
      \includegraphics[width=12 cm]{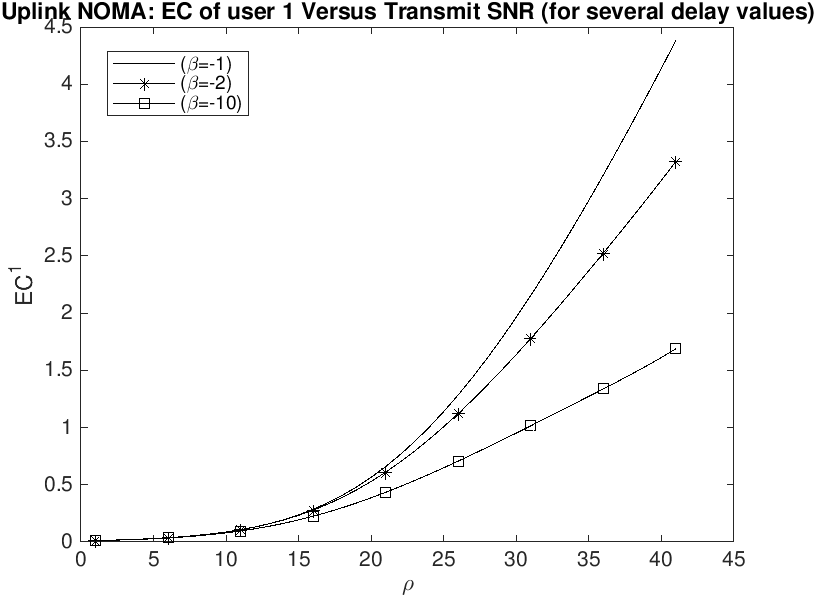}
      \caption{Effective capacity of user 1 $E_c^1$ in a two-user NOMA uplink network versus the transmit SNR $\rho$ (-10 to 30 dB) for several values of the normalized delay.}
      \label{fig7}
       \end{center}
  \end{figure}

\begin{figure}[ht]
       \begin{center}
      \includegraphics[width=12 cm]{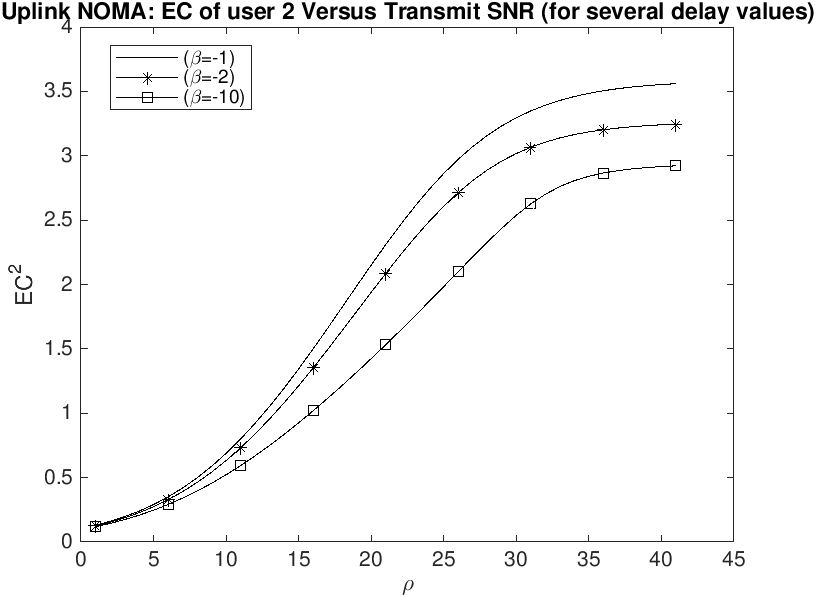}
      \caption{Effective capacity of user 2 $E_c^2$ in a two-user NOMA uplink network versus the transmit SNR $\rho$ (-10 to 30 dB) for several values of the normalized delay.}
      \label{fig8}
       \end{center}
  \end{figure}

 Figure \ref{fig6} includes the plots for the EC of the uplink two-user NOMA and OMA system versus the transmit SNR. We note that for the weak user, OMA is better than NOMA for low transmit SNRs, and NOMA is better than OMA at high transmit SNRs. In the contrary, for the strong user, NOMA is better than OMA at small values of SNR, and OMA is better at high values of the transmit SNR. We notice also that the EC of the strong is limited, it reaches a plateau at high SNR. This provides numerical validation of Lemma 1. \newline
 Figures \ref{fig7} and \ref{fig8}, show respectively the EC of user 1 and user 2, versus the transmit SNR, for several value of delay. when the delay become more stringent, i.e $\beta$ decreasing ( delay QoS increase), the individual link-layer rates in NOMA decreases, for the both users.


\begin{figure}[ht]
  \begin{center}
      \includegraphics[width=12 cm]{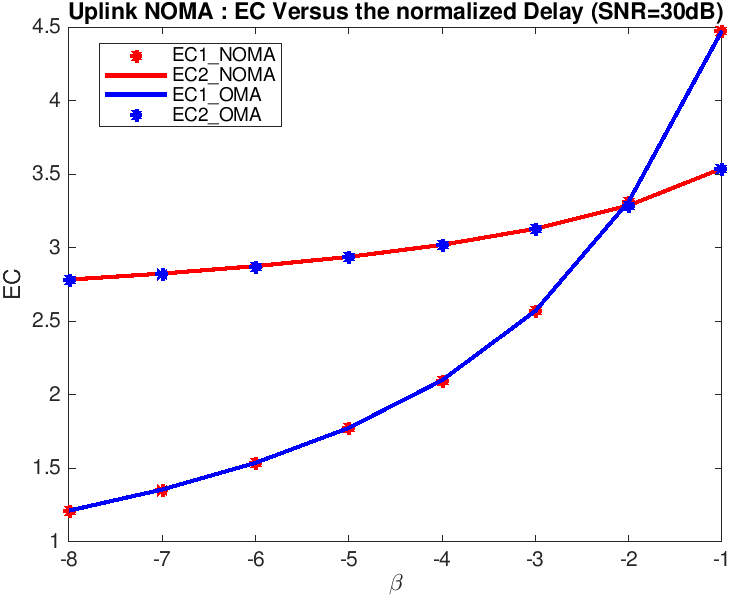}
      \caption{Effective capacity of user 1 and user 2 in a two-user NOMA uplink network compared to Ecs of two users OMA, versus the Normalized delay $\beta$ at $\rho$=30dB}
      \label{fig9}
     \end{center}
  \end{figure}

In Figure \ref{fig9}, the ECs of the strong and weak users are depicted as functions of the delay exponent, for NOMA and OMA scenarios. We note that EC in uplink NOMA and that in OMA are identical for both user.

Figure \ref{fig11} shows the difference of the EC in NOMA and the EC in OMA of the weak user. This curve starts initially at zero, then decreases to a certain minimum and starts increasing at high values of the transmit SNR. This confirms the Lemma 2. When the delay is equal to -1, we see that between 0 and 30 dB, the curves are negative, that means OMA outperform NOMA in this range. But After 30dB, the curves is positive, which means that NOMA offers better link-layer rate. However, that range O-30dB (where OMA outperforms NOMA) is not a general thing, we notice that by changing the transmission powers of users that range also changes. By increasing the transmission power of the weak user and reducing the transmission power of the strong user, we notice that the range is reduced. That range expands when we do the inverse. Also, when the delay becomes a little bit more stringent, $\beta_1$=$\beta_2$=-2, the zero crossing moves from 30 to 36 dB.

    \begin{figure}[ht]
     \begin{center}
      \includegraphics[width=12 cm]{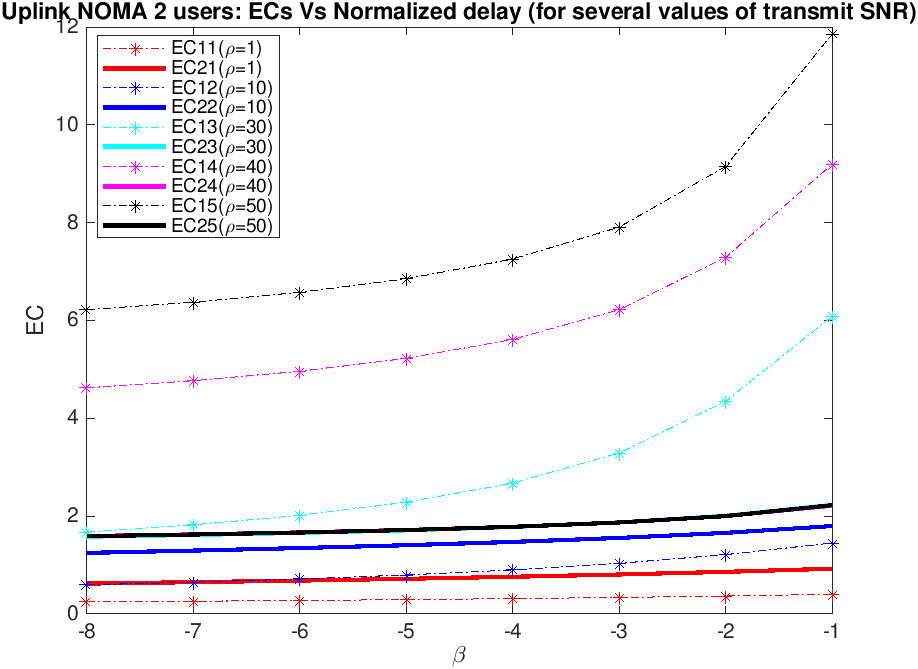}
      \caption{Effective capacities $E_c^1$ and $E_c^2$ in a two-user NOMA uplink network compared to Ecs of two users OMA, as functions of the Normalized delay $\beta$ (-8 to -1) for several values of the transmit SNR.}
      \label{fig10}
     \end{center}
  \end{figure}

      \begin{figure}[th]
       \begin{center}
      \includegraphics[width=12cm]{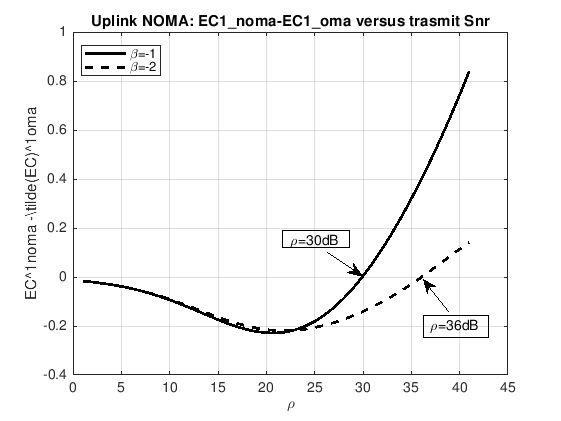}
      \caption{ $E_c^1-\Tilde{E}_c^1$ versus $\rho$ for various normalized delay.}
      \label{fig11}
     \end{center}
  \end{figure}

\begin{figure}[th]
  \begin{center}
      \includegraphics[width=12cm]{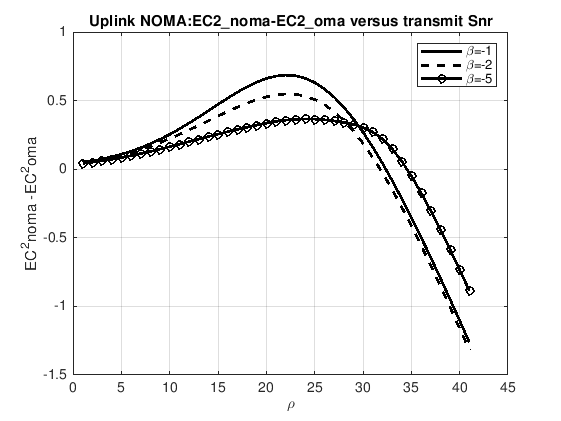}
      \caption{ $E_c^2-\Tilde{E}_c^2$ versus $\rho$ for various normalized delay.}
      \label{fig12}
   \end{center}
  \end{figure}

Figure \ref{fig12} shows the difference of the EC in NOMA and the EC in OMA of the strong user. This curve starts initially at zero, then increases to a certain maximum and starts decreasing to infinity at high values of the transmit SNR. This confirms Lemma 3. We note that the maximum of these curves decreases when the delay becomes more stringent.

\begin{figure}[h]
   \begin{center}
      \includegraphics[width=12cm]{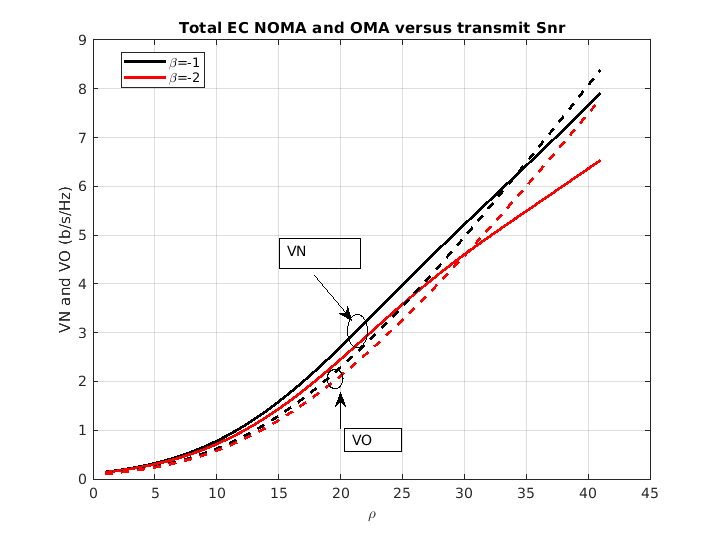}
      \caption{ $V_N$ and $V_O$ versus $\rho$ for various normalized delay.}
      \label{fig13}
    \end{center}
  \end{figure}

 To investigate the impact of $\rho$ on the performance of the total link-layer rate for the two-user system, Figure \ref{fig13} includes the plots for $V_N$ in NOMA and $V_O$ in OMA, versus the transmit SNR, for various delay. This figure shows that for both NOMA and OMA, the total EC for the two users, starts at the initial value of 0 and starts increasing with the transmit SNR. This confirms the Lemma 4.a and Lemma 4.d. This figure shows that when $\rho$ is very small, the total link-layer rate for the two user in NOMA, $V_N$, has a faster increasing slope than $V_O$ in OMA. On the contrary, with the increase of the transmit SNR, $V_O$ becomes gradually higher than $V_N$.  Further, at very high values of the SNR, the gap of the total EC between NOMA and OMA becomes unstable. And when the delay becomes more stringent, the total EC of both NOMA and OMA decreases.

\begin{figure}[h]
     \begin{center}
      \includegraphics[width=12cm]{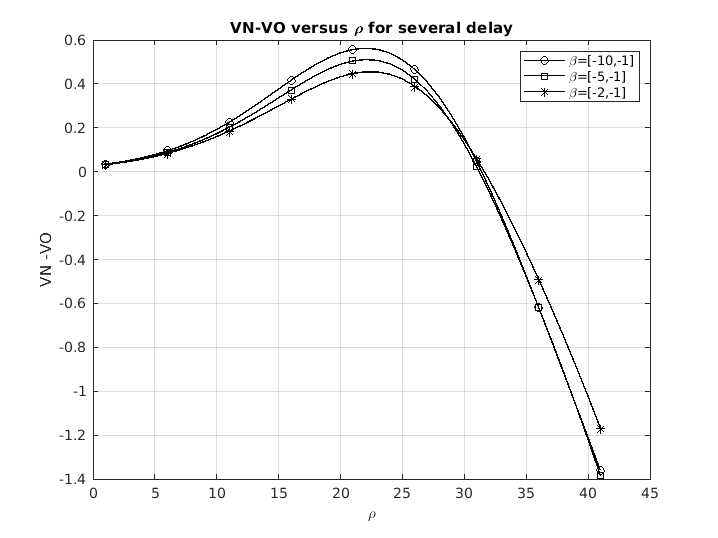}
      \caption{ $V_N$ - $V_O$ versus $\rho$ for various normalized delay.}
      \label{fig14}
      \end{center}
  \end{figure}

\begin{figure}[h]
   \begin{center}
      \includegraphics[width=12cm]{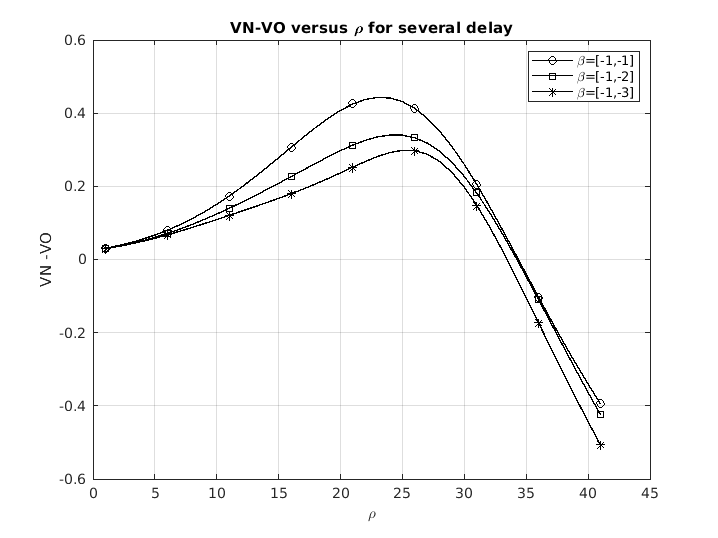}
      \caption{ $V_N$ - $V_O$ versus $\rho$ for various normalized delay.}
      \label{fig15}
     \end{center}
  \end{figure}

  Figures \ref{fig14} and \ref{fig15} depict the plots of EC versus $\rho$, for several normalized delay values. In figure \ref{fig14}, the delay of the strong user is fixed. And we see that when delay of the strong user is fixed, the highest delay QoS (i.e smallest normalized delay $\beta$) of the weak user gives us the highest level of the $V_N-V_O$.  On the other hand, when the delay is fixed, Figure \ref{fig15} shows that, the smallest delay Qos (i.e highest normalized delay $\beta$) value of the strong user gives us the highest level of the $V_N-V_O$.

  The curve of $V_N-V_O$ starts at zero,  increases to a maximum, and returns to zero. The transition to zero  is at $\rho$ = 31, and $\rho$ =36 respectively for the figures \ref{fig14} and \ref{fig15}. That means from 0 to 31dB (36dB in the Figure \ref{fig15}), the total link-layer rate of NOMA is higher than the OMA one. And After this transition to zero, the total link-layer rate OMA outperforms the NOMA one.

   \begin{figure}[h]
    \begin{center}
      \includegraphics[width=12cm]{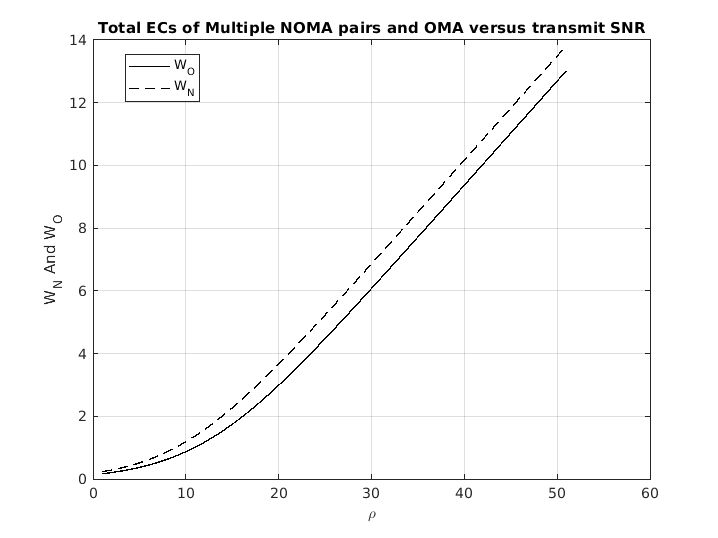}
      \caption{ $W_N$ and $W_O$ versus $\rho$}
      \label{fig16}
     \end{center}
  \end{figure}

 Then we focus on the comparison of  multiple NOMA pairs and OMA in term of the total link-layer rate, i.e $W_N$-$W_O$. The figure \ref{fig16} depicts the curve of the total link-layer of NOMA pairs, $W_N$, and the total link-layer of the OMA, versus the transmit SNR. The multiple NOMA pairs scheme outperforms the OMA. The performance gain of NOMA multiple users pairs over OMA start at zero, increases at small values of SNR, and becomes steady at high transmit SNRs.

    \begin{figure}[h]
     \begin{center}
      \includegraphics[width=12 cm]{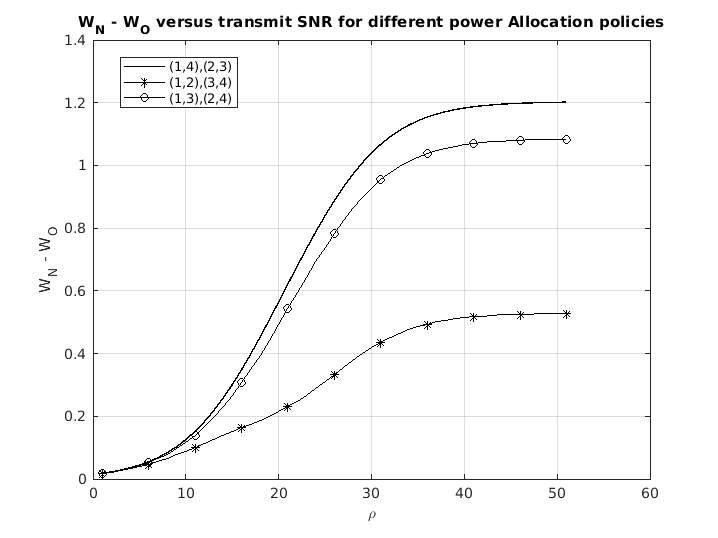}
      \caption{ $W_N$ - $W_O$ versus $\rho$ for various setting of user pairing set.}
      \label{fig17}
      \end{center}
  \end{figure}

 Figure \ref{fig17} shows the curves $W_N-W_O$ versus the transmit SNR, for various setting of the user-pairing. It starts at zero at small value of $\rho$, increases until it reaches a maximum, at high values of $\rho$. This Confirmes Lemma 6.  Specifically, we set the total number of users M=4, the power coefficients allocated to both users in a NOMA pair are given as P1=0,2 and P2=0.8 in all the groups and the normalized delay of all users are assumed to be equal $\beta_{1,i}$=$\beta_{2,i}$=-1, ($i=1,2,..,\frac{M}{2}$).

 \clearpage
\section{Conclusions and and Future Work}

The concept of the effective capacity enabled us to study the achievable data link layer rates when QoS delay guarantees are in place in the form of delay exponents. In this report, we have investigated the effective capacity of the uplink of a two-user NOMA network, assuming a block Rayleigh fading channel. We derived novel closed-form expressions for the two users in a two user-uplink NOMA network. We validated the proposed mathematical formulas with Monte Carlo simulations. We provided a comparison between a uplink NOMA two-user network and OMA uplink two-user network. We also analysed the impact of user-pairing on the total link, we showed that when pairing is used the total link layer rate is higher than the OMA one, and we found a best power allocation policy in the case four users, i.e., we found the set of pairs which gives the highest total link layer rate.
We showed that the ECs of both users decrease as the delay constraints become stringer. On the other hand, at high transmit SNRs, the EC of the weak user can surpass the EC of the strong user, as the latter is  limited due to interference.

In the uplink NOMA, the strong user decodes the message of all weaker users, before its own message. This is a problem from a security point of view. An eavesdropper can exploit this flaw to intercept communications. The next challenge is to find the maximum achievable link layer that can be achieved while guaranteeing secure communications. For that, another metric that has recently been defined can be used, the effective secrecy capacity (ESC) \cite{Wenjuan_secrecy}, \cite{qiao2010secure},\cite{hou2014effective}.  A performance analysis using that metric could be subject to future work, in the uplink NOMA scenario.

\underline{\textbf{Appendix A}}
\newline
\newline
From \eqref{EC1} we have:
\begin{equation}
  E_c^1 = \frac{1}{\beta_1}\log_2\left(2 \int_{0}^{\infty}(1+ \rho P_1 x_1)^{\beta_1} e^{-2x_1} dx_1\right)
\end{equation}
Set $t=\rho P_1 x_1$ i.e., $x_1$=$\frac{t}{\rho P_1}$ and since $x_1$:0$\rightarrow\infty$, $t$:0$\rightarrow\infty$, $dx_1$=$\frac{1}{\rho P_1}dt$.

Replacing these in (25):
\begin{equation}
  E_c^1 = \frac{1}{\beta_1}\log_2\left(\frac{2}{P_1 \rho} \int_{0}^{\infty}(1+ t)^{\beta_1} e^{-\frac{2t}{P_1\rho}}) dt\right)\nonumber
\end{equation}
setting, $a=1, (b-a-1)=\beta_1$, i.e.,  $b=\beta_2+2$. $z=\frac{2}{P_1 \rho}$ and having the $U\left(.,.,.\right)$ the confluent hypergeometric function U(a,b,z) defined as follow:\newline
$ U(a,b,z)=\frac{1}{\Gamma(a)}\int_0^{\infty}e^{-z t} t^{a-1}(1+t)^{b-a-1} dt $ \newline
We have:
\begin{equation}
 \int_{0}^{\infty}(1+ t)^{\beta_1} e^{-\frac{2t}{P_1\rho}} dt=U(1,2+\beta_1,\frac{2}{\rho P_1})
\end{equation}
Finally we have the closed-form expression of the User 1:

For the second user, by replacing from \eqref{EC2} have:
\begin{eqnarray}
  E_c^2 &=& \frac{1}{\beta_2}\log_2(\mathbf{E}[(1+ \frac{\rho P_2 x_{2}}{1+ \rho P_1 x_{1}})^{\beta_2}])\\
  &=&\frac{1}{\beta_2}\log_2\Big(\int_{0}^{\infty}\int_{x_{1}}^{\infty}\Big(1+ \frac{\rho P_2 x_2}{1+\rho P_1 x_1}\Big)^{\beta_2}f_{\gamma_{1:2}}(x_1,x_2)dx_2 dx_1\Big)\\
  &=&\frac{1}{\beta_2}\log_2\Big(2\int_{0}^{\infty}\int_{x_{1}}^{\infty}\Big(1+ \frac{\rho P_2 x_2}{1+\rho P_1 x_1}\Big)^{\beta_2}e^{-x_1}e^{-x_2}dx_2 dx_1\Big)\\
  &=&\frac{1}{\beta_2}\log_2\Big(2\int_{0}^{\infty}(\frac{\rho P_2}{1+\rho P_1 x_1})^{\beta_2}e^{-x_1}\int_{x_{1}}^{\infty}\Big(\frac{1+\rho P_1 x_1}{\rho P_2}+ x_2\Big)^{\beta_2}e^{-x_2}dx_2 dx_1 \Big) \nonumber
  \label{eq4}
\end{eqnarray}

Setting $z=\frac{1+\rho P_1 x_1}{\rho P_2}+ x_2$, we have: $x_2=z-\frac{1+\rho P_1 x_1}{\rho P_2}$ and $dx_2=dz$, so that
when $x_2\rightarrow x_1$, $z \rightarrow \frac{1+\rho P_1 x_1}{\rho P_2}+ x_1=\frac{1+\rho x_1}{\rho P_2}$ and when $x_2 \rightarrow \infty$, $z \rightarrow \infty$
\begin{eqnarray}
  E_c^2 &=&\frac{1}{\beta_2}\log_2\Big(2\int_{0}^{\infty}(\frac{\rho P_2}{1+\rho P_1 x_1})^{\beta_2}e^{-x_1}\int_{\frac{1+\rho x_1}{\rho P_2}}^{\infty}z^{\beta_2}e^{-(z-\frac{1+\rho P_1 x_1}{\rho P_2})}dz dx_1 \Big)\\
  &=&\frac{1}{\beta_2}\log_2\Big(2e^{\frac{1}{\rho P_2}}\int_{0}^{\infty}(\frac{\rho P_2}{1+\rho P_1 x_1})^{\beta_2}e^{-x_1}e^{\frac{P_1 x_1}{ P_2}}\int_{\frac{1+\rho x_1}{\rho P_2}}^{\infty}z^{\beta_2}e^{-z}dz dx_1 \Big)\nonumber
  \label{eq4}
\end{eqnarray}
we note that
$ \int_{a}^{\infty}\frac{e^{-x}}{x^b}dx=a^{-\frac{b}{2}} e^{-\frac{a}{2}}\mathbf{W}_{-\frac{b}{2},-\frac{1-b}{2}}(a)$
, where $\mathbf{W}$ is the Whittaker W function.
i.e.,
\begin{eqnarray}
  E_c^2 &=&\frac{1}{\beta_2}\log_2\Big(2e^{\frac{1}{\rho P_2}}\int_{0}^{\infty}(\frac{\rho P_2}{1+\rho P_1 x_1})^{\beta_2}e^{-x_1}e^{\frac{P_1 x_1}{ P_2}}\Big[(\frac{1+\rho x_1}{\rho P_2})^{\frac{\beta_2}{2}}e^{-\frac{1+\rho x_1}{2 \rho P_2}}\mathbf{W}_{\frac{\beta_2}{2},\frac{1+\beta_2}{2}}(\frac{1+\rho x_1}{\rho P_2}) \Big] dx_1 \Big)\nonumber\\
  &=&\frac{1}{\beta_2}\log_2\Big(2e^{\frac{1}{2 \rho P_2}}\int_{0}^{\infty}(\rho P_2)^{\frac{\beta_2}{2}}(1+\rho P_1 x_1)^{-\beta_2} (1+\rho x_1)^{\frac{\beta_2}{2}}e^{-x_1}e^{\frac{P_1 x_1}{ P_2}}e^{-\frac{x_1}{2 P_2}}\Big[\mathbf{W}_{\frac{\beta_2}{2},\frac{1+\beta_2}{2}}(\frac{1+\rho x_1}{\rho P_2}) \Big] dx_1 \Big)\nonumber\\
  &=&\frac{1}{\beta_2}\log_2\Big(2(\rho P_2)^{\frac{\beta_2}{2}} e^{\frac{1}{2 \rho P_2}}\int_{0}^{\infty}(1+\rho P_1 x_1)^{-\beta_2} (1+\rho x_1)^{\frac{\beta_2}{2}}e^{\frac{(2P_1-2 P_2-1) x_1}{2 P_2}}\Big[\mathbf{W}_{\frac{\beta_2}{2},\frac{1+\beta_2}{2}}(\frac{1+\rho x_1}{\rho P_2}) \Big] dx_1 \Big)
  \nonumber
  \label{eq4}
\end{eqnarray}
Note that
$\mathbf{W}_{u-\frac{1}{2},u}(z)=e^{\frac{1}{2}z} z^{\frac{1}{2}-u}\Gamma(2 u,z)$, so that we have
$\mathbf{W}_{\frac{\beta_2}{2},\frac{1+\beta_2}{2}}(\frac{1+\rho x_1}{\rho P_2})=e^{\frac{1+\rho x_1}{2 \rho P_2}}(\frac{1+\rho x_1}{\rho P_2})^{-\frac{\beta_2}{2}}\Gamma(1+\beta_2,\frac{1+\rho x_1}{\rho P_2})$.
Replacing that in EC, we have: \newline
\begin{align}
  E_c^2 &=\frac{1}{\beta_2}\log_2\Bigg(2(\rho P_2)^{\frac{\beta_2}{2}} e^{\frac{1}{2 \rho P_2}}\int_{0}^{\infty}(1+\rho P_1 x_1)^{-\beta_2} (1+\rho x_1)^{\frac{\beta_2}{2}}e^{\frac{(2P_1-2 P_2-1) x_1}{2 P_2}}\Big[e^{\frac{1+\rho x_1}{2 \rho P_2}}(\frac{1+\rho x_1}{\rho P_2})^{-\frac{\beta_2}{2}}\nonumber \\
  &\times\Gamma(1+\beta_2,\frac{1+\rho x_1}{\rho P_2}) \Big] dx_1 \Bigg)\\
  &=\frac{1}{\beta_2}\log_2\Bigg(2(\rho P_2)^{\frac{\beta_2}{2}} e^{\frac{1}{2 \rho P_2}}\int_{0}^{\infty}(1+\rho P_1 x_1)^{-\beta_2} (1+\rho x_1)^{\frac{\beta_2}{2}}e^{\frac{(2P_1-2 P_2-1) x_1}{2 P_2}}\Big[e^{\frac{1}{2 \rho P_2}}e^{\frac{x_1}{2 P_2}}(1+\rho x_1)^{-\frac{\beta_2}{2}}(\rho P_2)^{\frac{\beta_2}{2}})\nonumber \\
  &\times\Gamma(1+\beta_2,\frac{1+\rho x_1}{\rho P_2}) \Big] dx_1 \Bigg)\nonumber\\
  &=\frac{1}{\beta_2}\log_2\Bigg(2(\rho P_2)^{\beta_2} e^{\frac{1}{\rho P_2}}\int_{0}^{\infty}(1+\rho P_1 x_1)^{-\beta_2} e^{\frac{(P_1-P_2) x_1}{P_2}}\Big[\Gamma(1+\beta_2,\frac{1+\rho x_1}{\rho P_2}) \Big] dx_1 \Bigg)
  \nonumber
  \label{eq4}
\end{align}
Set $\frac{1+\rho x_1}{\rho P_2}=y$, i.e $x_1=P_2 y-\frac{1}{\rho}$, and $dx_1=P_2dy$. When $x_1\rightarrow0$, $y\rightarrow\frac{1}{\rho P_2}$. 
When $x_1\rightarrow\infty$, $y\rightarrow\infty$
while we also have $P_1+P_2=1$.
\begin{align}
  E_c^2 &=\frac{1}{\beta_2}\log_2\Bigg(2(\rho P_2)^{\beta_2} e^{\frac{1}{\rho P_2}}\int_{0}^{\infty}(1+\rho P_1 x_1)^{-\beta_2} e^{\frac{(P_1-P_2) x_1}{P_2}}\Big[\Gamma(1+\beta_2,\frac{1+\rho x_1}{\rho P_2}) \Big] dx_1 \Bigg)\nonumber\\
  &=\frac{1}{\beta_2}\log_2\Bigg(2(\rho P_2)^{\beta_2} e^{\frac{1}{\rho P_2}}\int_{\frac{1}{\rho P_2}}^{\infty}(1+\rho P_1 (P_2 y-\frac{1}{\rho}))^{-\beta_2}e^{\frac{(P_1-P_2) (P_2 y-\frac{1}{\rho})}{P_2}}\Big[\Gamma(1+\beta_2,y) \Bigg] P_2 dy \Bigg)\nonumber\\
  &=\frac{1}{\beta_2}\log_2\Bigg(2 P_2 (\rho P_2)^{\beta_2} e^{\frac{1}{\rho P_2}}e^{-\frac{(P_1-P_2)}{\rho P_2}}\times\int_{\frac{1}{\rho P_2}}^{\infty}(1-P_1+\rho P_1P_2 y)^{-\beta_2} e^{(P_1-P_2)y}\Gamma(1+\beta_2,y) dy \Big)\nonumber\\
 & =\frac{1}{\beta_2}\log_2\Bigg(2 P_2 (\rho P_2)^{\beta_2} e^{\frac{1}{\rho P_2}}e^{-\frac{(P_1-P_2)}{\rho P_2}}\times\int_{\frac{1}{\rho P_2}}^{\infty}P_2^{-\beta_2}(1+\rho P_1 y)^{-\beta_2} e^{(P_1-P_2)y}\Gamma(1+\beta_2,y) dy \Bigg)
  \nonumber
\end{align}

Using the binomial expansion we have
$(1+\rho P_1 y)^{-\beta_2}=\sum_{j=0}^{-\beta_2}\binom{-\beta_2}{j}(\rho P_1 y)^{j}$
and using Taylor series expansion we have
 $ e^{(P_1-P_2)y}= e^{-(P_2-P_1)y}=\sum_{k=0}^{\infty}\frac{(-1)^k (P_2-P_1)^k}{k!}y^k$, which converges.

\begin{align}
  E_c^2 &
  =\frac{1}{\beta_2}\log_2\Bigg(2 P_2^{1-\beta_2}(\rho P_2)^{\beta_2} e^{\frac{1}{\rho P_2}}e^{-\frac{(P_1-P_2)}{\rho P_2}}\times\int_{\frac{1}{\rho P_2}}^{\infty}(1+\rho P_1 y)^{-\beta_2} e^{(P_1-P_2)y}\Gamma(1+\beta_2,y) dy \Bigg)\nonumber\\
 & =\frac{1}{\beta_2}\log_2\Bigg(2 P_2^{1-\beta_2}(\rho P_2)^{\beta_2} e^{\frac{1}{\rho P_2}}e^{-\frac{(P_1-P_2)}{\rho P_2}}\nonumber\\
 &\times\int_{\frac{1}{\rho P_2}}^{\infty}\sum_{j=0}^{-\beta_2}\binom{-\beta_2}{j}(\rho P_1 y)^{j}\times \sum_{k=0}^{\infty}\frac{(-1)^k (P_2-P_1)^k}{k!}y^k \times \Gamma(1+\beta_2,y) dy \Bigg)\nonumber\\
  &=\frac{1}{\beta_2}\log_2\Bigg(2 P_2^{1-\beta_2}(\rho P_2)^{\beta_2} e^{\frac{1}{\rho P_2}}e^{-\frac{(P_1-P_2)}{\rho P_2}}\times\sum_{j=0}^{-\beta_2}\binom{-\beta_2}{j}(\rho P_1)^{j}
  \sum_{k=0}^{\infty}\frac{(-1)^k (P_2-P_1)^k}{k!}\int_{\frac{1}{\rho P_2}}^{\infty}y^{j+k} \Gamma(1+\beta_2,y) dy \Bigg).
  \nonumber
\end{align}

Note that
\begin{equation}
  \int_{c}^{\infty}y^{b} \Gamma(A,z) dz  =\frac{1}{1+b}\Big(-c^{1+b}\Gamma[A,c]+\Gamma[1+A+b,c]\Big)\nonumber
\end{equation}
i.e.,
\begin{align}
\int_{\frac{1}{\rho P_2}}^{\infty}y^{j+k} \Gamma(1+\beta_2,y) dy=
\frac{1}{1+j+k}\Big[-(\rho P_2)^{-1-j-k}\Gamma[1+\beta_2,\frac{1}{\rho P_2}]+\Gamma[2+\beta_2+j+k,\frac{1}{\rho P_2}]\Big]
\end{align}
Then finally we get the closed form expression for the strong User:
\begin{align}
  E_c^2=&
  \frac{1}{\beta_2}\log_2\Big(2 P_2^{1-\beta_2}(\rho P_2)^{\beta_2} e^{\frac{1}{\rho P_2}}e^{-\frac{(P_1-P_2)}{\rho P_2}}\Big)\nonumber \\ &+\frac{1}{\beta_2}\log_2\Big(\sum_{j=0}^{-\beta_2}\binom{-\beta_2}{j}(\rho P_1)^{j}\times \sum_{k=0}^{\infty}\frac{(-1)^k (P_2-P_1)^k}{k!}\frac{1}{1+j+k}\nonumber \\
  &\times\Big[\Gamma[2+\beta_2+j+k,\frac{1}{\rho P_2}]-(\rho P_2)^{-1-j-k}\Gamma[1+\beta_2,\frac{1}{\rho P_2}] \Big] \Big)
  \nonumber
  \label{eq4}
\end{align}

\underline{\textbf{Appendix B}}

The closed-form expression for the EC OMA, of the mth user with M total users, is determined in \cite{yu2018link} as follows:
\begin{align}
\widetilde{E}_c^{m}&=\frac{1}{\beta_m}\log_2\Big(E\Big[(1+\rho |h_m|^2)^{\frac{\beta_m}{2}}\Big] \Big)=\frac{1}{\beta_m}\log_2\Bigg(\frac{\psi_m}{\rho} \int_{0}^{\infty}(1+ \gamma_m)^{\beta_m} e^{-\frac{(M-m+1)\gamma_m}{\rho}}\Bigg)
\nonumber \\
&=\frac{1}{\beta_m}\log_2\Big(\frac{\psi_m}{\rho}\sum_{k=0}^{m-1}\binom{m-1}{k}(-1)^{k} \int_{0}^{\infty}(1+ \gamma_m)^{\beta_m}e^{-\frac{(M-m+1+k)\gamma_m}{\rho}}d\gamma_m \Big)\nonumber
\end{align}
so that
\begin{equation}
\widetilde{E}_c^{m}=\frac{1}{\beta_m}\log_2\Big(\frac{\psi_m}{\rho}\sum_{k=0}^{m-1}\binom{m-1}{k}(-1)^{k} \\
\times U\left(1,2+\frac{2}{M}\beta_m,\frac{M-m+1+k}{\rho}\right))
\end{equation}

For the two users case, $M=2$, we have:
\begin{equation}
\widetilde{E}_c^{1}=\frac{1}{\beta_1}\log_2\Big(\frac{2}{\rho}
\times U\left(1,2+\beta_1,\frac{2}{\rho}\right)\Big)
\end{equation}
and, 
\begin{equation}
\widetilde{E}_c^{2}=\frac{1}{\beta_2}\log_2\Big(\frac{2}{\rho}\sum_{k=0}^{1}\binom{1}{k}(-1)^{k}
\times U\left(1,2+\beta_2,\frac{1+k}{\rho}\right))
\end{equation}
Where $U(\cdot)$ is the confluent hypergeometric function of the second kind, defined as follow:
\begin{equation}
    U(a,b,z)=\frac{1}{\Gamma(a)}\int_{0}^{\infty}e^{-z t} t^{a-1}(1+t)^{b-a-1} dt,\\ \text{for }Re( a) , Re( z) > 0,
\end{equation}

\underline{\textbf{Appendix C:}}

Here the proof of Lemma 1:
By inserting $\rho\rightarrow0$, in (9),(10), (13), (14), $E_c^1-\widetilde{E}_c^{1}$ and $E_c^2-\widetilde{E}_c^{2}$ we get the (a) in the lemma 1.

\begin{equation}
   \lim\limits_{\rho\rightarrow0}(E_c^1-\widetilde{E}_c^{1})=\frac{1}{\beta_1}\log_2(\frac{\mathbb{E}[(1+\rho P_1 |h_1|^2)^{\beta_2}]}{\mathbb{E}[(1+\rho |h_1|^2)^{\frac{\beta_2}{2}}]})=0 \nonumber
\end{equation}
\begin{equation}
   \lim\limits_{\rho\rightarrow0}(E_c^2-\widetilde{E}_c^{2})=\frac{1}{\beta_2}\log_2(\frac{\mathbb{E}[(1+\frac{\rho P_2 |h_2|^2}{1+\rho P_1 |h_1|^2})^{\beta_2}]}{\mathbb{E}[(1+\rho |h_1|^2)^{\frac{\beta_2}{2}}]})=0 \nonumber
\end{equation}

In the same way, by inserting $\rho\rightarrow \infty$, in (9),(10), (13), (14), $E_c^1-\widetilde{E}_c^{1}$ and $E_c^2-\widetilde{E}_c^{2}$ we get the lemma 1.(b)

\begin{equation}
\lim\limits_{\rho\rightarrow \infty}E_c^2\rightarrow \frac{1}{\beta_2}\log_2(\mathbb{E}[(1+\frac{P_2 |h_2|^2}{P_1 |h_1|^2})^{\beta_2}]) \nonumber
\end{equation}

\begin{equation}
   \lim\limits_{\rho\rightarrow \infty}(E_c^1-\widetilde{E}_c^{1})=\frac{1}{\beta_1}\log_2(\rho^{\frac{\beta_1}{2}} \frac{\mathbb{E}[(\frac{1}{\rho}+P_1 |h_1|^2)^{\beta_2}]}{\mathbb{E}[(\frac{1}{\rho}+ |h_1|^2)^{\frac{\beta_2}{2}}]})=\infty \nonumber
\end{equation}{}
\begin{equation}
   \lim\limits_{\rho\rightarrow \infty}(E_c^2-\widetilde{E}_c^{2})=\frac{1}{\beta_2}\log_2(\frac{\mathbb{E}[(\frac{\frac{1}{\rho}+ P_1 |h_1|^2+ P_2 |h_2|^2}{\frac{1}{\rho}+ P_1 |h_1|^2})^{\beta_2}]}{\rho^{\frac{\beta_2}{2}}\mathbb{E}[(\frac{1}{\rho}+ |h_1|^2)^{\frac{\beta_2}{2}}]})=-\infty \nonumber
\end{equation}{}

\underline{\textbf{Appendix D:}}

To analyze the trends of $E_c^1$ and $\widetilde{E}_c^{1}$ with respect to $\rho$ :
\begin{equation}
    \frac{\partial E_c^1}{\partial\rho}=\frac{1}{\beta_1 \ln2}\frac{\Big(\mathbb{E}[(1+\rho P_1 |h_1|^2)^{\beta_1}] \Big)'}{\mathbb{E}[(1+\rho P_1 |h_1|^2)^{\beta_1}]}=\frac{P_1}{\ln2} \frac{\mathbb{E}[|h_1|^2 (1+\rho P_1 |h_1|^2)^{\beta_1-1}]}{\mathbb{E}[(1+\rho P_1 |h_1|^2)^{\beta_1}]}\ge 0   \nonumber
\end{equation}

In the same way we get for user 1 OMA:
\begin{equation}
    \frac{\partial \widetilde{E}_c^{1}}{\partial\rho}=\frac{1}{\beta_1\ln2}\frac{\Big(\mathbb{E}[(1+\rho |h_1|^2)^{\frac{\beta_1}{2}}] \Big)'}{\mathbb{E}[(1+\rho |h_1|^2)^{\frac{\beta_1}{2}}]}=\frac{1}{2 \ln2} \frac{\mathbb{E}[|h_1|^2 (1+\rho |h_1|^2)^{\frac{\beta_1}{2}-1}]}{\mathbb{E}[(1+\rho |h_1|^2)^{\frac{\beta_1}{2}}]} \ge 0   \nonumber
\end{equation}

Then:
\begin{equation}
    \frac{\partial (E_c^1-\widetilde{E}_c^{1})}{\partial\rho}=\frac{P_1}{\ln2} \frac{\mathbb{E}[|h_1|^2 (1+\rho P_1 |h_1|^2)^{\beta_1-1}]}{\mathbb{E}[(1+\rho P_1 |h_1|^2)^{\beta_1}]}-\frac{1}{2 \ln2} \frac{\mathbb{E}[|h_1|^2 (1+\rho |h_1|^2)^{\frac{\beta_1}{2}-1}]}{\mathbb{E}[(1+\rho |h_1|^2)^{\frac{\beta_1}{2}}]}\nonumber
\end{equation}

$\lim\limits_{\rho\rightarrow0}(\frac{\partial (E_c^1-\widetilde{E}_c^{1})}{\partial\rho})=\frac{(P_1-\frac{1}{2})}{\ln 2}\mathbb{E}[|h_1|^2] \leq0$,

When $\rho$ is very large,
\begin{equation}
\frac{\partial (E_c^1-\widetilde{E}_c^{1})}{\partial\rho})=\frac{P_1}{\ln2} \frac{\mathbb{E}[|h_1|^2 (\rho P_1 |h_1|^2)^{\beta_1-1}]}{\mathbb{E}[(\rho P_1 |h_1|^2)^{\beta_1}]}-\frac{1}{2 \ln2} \frac{\mathbb{E}[|h_1|^2 (\rho |h_1|^2)^{\frac{\beta_1}{2}-1}]}{\mathbb{E}[(\rho |h_1|^2)^{\frac{\beta_1}{2}}]}=
\frac{1}{2\rho \ln2}\ge0 \nonumber
\end{equation}
when $\rho\rightarrow\infty$, this term approaches 0.

\underline{\textbf{Appendix E:}}

\begin{equation}
E_c^2=\frac{1}{\beta_2}\log_2(\mathbf{E}[(1+ \frac{\rho P_2 |h_2|^2}{1+  \rho P_1 |h_1|^2})^{\beta_2}])\nonumber
\end{equation}

\begin{equation}
    \frac{\partial E_c^2}{\partial\rho}=\frac{1}{\beta_2 \ln2}\frac{\Big(\mathbf{E}[(1+ \frac{\rho P_2 |h_2|^2}{1+  \rho P_1 |h_1|^2})^{\beta_2}]\Big)'}{\mathbf{E}[(1+ \frac{\rho P_2 |h_2|^2}{1+  \rho P_1 |h_1|^2})^{\beta_2}]}=\frac{1}{ \ln2}\frac{\mathbf{E}[\frac{P_2 |h_2|^2}{(1+\rho P_1 |h_1|^2)^2}(1+ \frac{\rho P_2 |h_2|^2}{1+  \rho P_1 |h_1|^2})^{\beta_2-1}]}{\mathbf{E}[(1+ \frac{\rho P_2 |h_2|^2}{1+  \rho P_1 |h_1|^2})^{\beta_2}]}\ge0 \nonumber
\end{equation} 
In the same way, for EC in OMA for the user 2:
\begin{equation}
    \frac{\partial \widetilde{E}_c^{2}}{\partial\rho}=\frac{1}{\beta_2\ln2}\frac{\Big(\mathbb{E}[(1+\rho |h_2|^2)^{\frac{\beta_2}{2}}] \Big)'}{\mathbb{E}[(1+\rho |h_2|^2)^{\frac{\beta_2}{2}}]}\\=\frac{1}{2 \ln2} \frac{\mathbb{E}[|h_2|^2 (1+\rho |h_2|^2)^{\frac{\beta_2}{2}-1}]}{\mathbb{E}[(1+\rho |h_2|^2)^{\frac{\beta_2}{2}}]} \ge 0   \nonumber
\end{equation}

and
\begin{equation}
\frac{\partial (E_c^2-\widetilde{E}_c^{2})}{\partial\rho}=\frac{1}{ \ln2}\frac{\mathbf{E}[\frac{P_2 |h_2|^2}{(1+\rho P_1 |h_1|^2)^2}(1+ \frac{\rho P_2 |h_2|^2}{1+  \rho P_1 |h_1|^2})^{\beta_2-1}]}{\mathbf{E}[(1+ \frac{\rho P_2 |h_2|^2}{1+  \rho P_1 |h_1|^2})^{\beta_2}]}\\ -\frac{1}{2 \ln2} \frac{\mathbb{E}[|h_2|^2 (1+\rho |h_2|^2)^{\frac{\beta_2}{2}-1}]}{\mathbb{E}[(1+\rho |h_2|^2)^{\frac{\beta_2}{2}}]} \nonumber
\end{equation}

When $\rho\rightarrow0$, 
$\lim\limits_{\rho\rightarrow0}(\frac{\partial (E_c^2-\widetilde{E}_c^{2})}{\partial\rho})=\frac{(P_2-\frac{1}{2})}{\ln 2}\mathbb{E}[|h_2|^2]$.

When $\rho$ is very large,
\begin{align}
\frac{\partial (E_c^2-\widetilde{E}_c^{2})}{\partial\rho}&=\frac{1}{ \ln2}\frac{\mathbf{E}[\frac{P_2 |h_2|^2}{\rho^2 (\frac{1}{\rho}+ P_1 |h_1|^2)^2}(1+\frac{\rho}{\rho} \frac{(P_2 |h_2|^2)}{(\frac{1}{\rho}+ P_1 |h_1|^2)})^{\beta_2-1}]}{\mathbf{E}[(1+ \frac{\rho}{\rho}\frac{P_2 |h_2|^2}{(\frac{1}{\rho}+ P_1 |h_1|^2)})^{\beta_2}]}
-\frac{1}{2 \ln2}\frac{1}{\rho} \frac{\mathbb{E}[|h_2|^2 (\frac{1}{\rho}+ |h_2|^2)^{\frac{\beta_2}{2}-1}]}{\mathbb{E}[(\frac{1}{\rho}+ |h_2|^2)^{\frac{\beta_2}{2}}]}\nonumber\\
&=\frac{1}{ \ln2}\frac{\mathbf{E}[\frac{P_2 |h_2|^2}{\rho^2 (P_1 |h_1|^2)^2}(1+\frac{P_2 |h_2|^2}{P_1 |h_1|^2})^{\beta_2-1}]}{\mathbf{E}[(1+\frac{P_2 |h_2|^2}{P_1 |h_1|^2})^{\beta_2}]}
-\frac{1}{2 \ln2}\frac{1}{\rho} \frac{\mathbb{E}[( |h_2|^2)^{\frac{\beta_2}{2}}]}{\mathbb{E}[( |h_2|^2)^{\frac{\beta_2}{2}}]}\nonumber\\
&=\frac{P_2}{\rho^2 P_{1}^2 \ln2}\frac{\mathbf{E}[\frac{|h_2|^2}{(|h_1|^2)^2}(1+\frac{P_2 |h_2|^2}{P_1 |h_1|^2})^{\beta_2-1}]}{\mathbf{E}[(1+\frac{P_2 |h_2|^2}{P_1 |h_1|^2})^{\beta_2}]} -\frac{1}{2 \ln2}\frac{1}{\rho} \nonumber\\
&=\frac{\frac{P_2}{P_{1}^2\ln2} A-\frac{1}{2 \ln2}\rho}{\rho^2}\nonumber
\end{align}
Where $A=\frac{\mathbf{E}[\frac{|h_2|^2}{(|h_1|^2)^2}(1+\frac{P_2 |h_2|^2}{P_1 |h_1|^2})^{\beta_2-1}]}{\mathbf{E}[(1+\frac{P_2 |h_2|^2}{P_1 |h_1|^2})^{\beta_2}]}$, unrelated to $\rho$. \newline
So when $\rho$ is very large, $\frac{\partial (E_c^2-\widetilde{E}_c^{2})}{\partial\rho}$ can be approximated by $-\frac{1}{2 \ln2}\frac{1}{\rho}$, and it gradually approaches 0 when $\rho\rightarrow\infty$.
\newline

\underline{\textbf{Appendix F:}}
\newline
Here is provided the proof of the Lemma 4.

$V_N=E_c^1+E_c^{2}$, using Lemma 1, we have 
$ \lim\limits_{\rho\rightarrow0}(V_N)=0$ and $ \lim\limits_{\rho\rightarrow\infty}(V_N)=\infty$.

\begin{equation}
\frac{\partial V_N}{\partial\rho}=\frac{\partial (E_c^1+E_c^{2})}{\partial\rho}\\=\frac{P_1}{\ln2} \frac{\mathbb{E}[|h_1|^2 (1+\rho P_1 |h_1|^2)^{\beta_1-1}]}{\mathbb{E}[(1+\rho P_1 |h_1|^2)^{\beta_1}]}+\\\frac{1}{ \ln2}\frac{\mathbf{E}[\frac{P_2 |h_2|^2}{(1+\rho P_1 |h_1|^2)^2}(1+ \frac{\rho P_2 |h_2|^2}{1+  \rho P_1 |h_1|^2})^{\beta_2-1}]}{\mathbf{E}[(1+ \frac{\rho P_2 |h_2|^2}{1+  \rho P_1 |h_1|^2})^{\beta_2}]},\nonumber
\end{equation}
which is non negative, because $ \frac{\partial E_c^1}{\partial\rho}\ge0$ and $ \frac{\partial E_c^2}{\partial\rho}\ge0$ as we previously showed.

When $\rho\rightarrow0$, we have:
$ \lim\limits_{\rho\rightarrow 0}(\frac{\partial V_N}{\partial\rho})=\frac{P_1}{\ln2}\mathbb{E}[|h_1|^2]+\frac{P_2}{ \ln2}\mathbb{E}[|h_2|^2]$
\newline

When $\rho\rightarrow\infty$,
\begin{equation}
 \lim\limits_{\rho\rightarrow\infty}(\frac{\partial V_N}{\partial\rho})=\frac{1}{\rho \ln2}+\frac{1}{\rho^2 \ln2}\frac{\mathbf{E}[\frac{P_2 |h_2|^2}{(P_1 |h_1|^2)^2}(1+ \frac{P_2 |h_2|^2}{P_1 |h_1|^2})^{\beta_2-1}]}{\mathbf{E}[(1+ \frac{P_2 |h_2|^2}{ P_1 |h_1|^2})^{\beta_2}]} \nonumber
\end{equation}
which equals to $0$.

Same thing is done for the OMA, for $V_O$:

$V_O=\widetilde{E}_c^{1}+\widetilde{E}_c^{2}$, using Lemma 1, we have 
$ \lim\limits_{\rho\rightarrow0}(V_0)=0$ and $ \lim\limits_{\rho\rightarrow\infty}(V_0)=\infty$.

\begin{equation}
\frac{\partial V_0}{\partial\rho}=\frac{\partial (\widetilde{E}_c^{1}+\widetilde{E}_c^{2})}{\partial\rho}\\=\frac{1}{2 \ln2} \frac{\mathbb{E}[|h_1|^2 (1+\rho |h_1|^2)^{\frac{\beta_1}{2}-1}]}{\mathbb{E}[(1+\rho |h_1|^2)^{\frac{\beta_1}{2}}]}+\\\frac{1}{2 \ln2} \frac{\mathbb{E}[|h_2|^2 (1+\rho |h_2|^2)^{\frac{\beta_2}{2}-1}]}{\mathbb{E}[(1+\rho |h_2|^2)^{\frac{\beta_2}{2}}]},\nonumber
\end{equation}
which is non negative, because $ \frac{\partial\widetilde{E}_c^{1}}{\partial\rho}\ge0$ and $ \frac{\partial \widetilde{E}_c^{2}}{\partial\rho}\ge0$ as we previously showed.

When $\rho\rightarrow0$, we have:
$ \lim\limits_{\rho\rightarrow 0}(\frac{\partial V_O}{\partial\rho})=\frac{1}{2 \ln2}\mathbb{E}[|h_1|^2]+\frac{1}{2  \ln2}\mathbb{E}[|h_2|^2]$

When $\rho\rightarrow\infty$,
\begin{equation}
 \lim\limits_{\rho\rightarrow\infty}(\frac{\partial V_O}{\partial\rho})= \lim\limits_{\rho\rightarrow\infty}(\frac{1}{2 \rho \ln2}+\frac{1}{2 \rho \ln2})=\lim\limits_{\rho\rightarrow\infty}(\frac{1}{ \rho \ln2}) \nonumber
\end{equation}
which equals to $0$.

\underline{\textbf{Appendix G:}}

We have:
\begin{align}
   E_c^2 &= \frac{1}{\beta_2}\log_2(\mathbb{E}[(1+ \frac{\rho P_2 |h_2|^2}{1+ \rho P_1 |h_1|^2})^{\beta_2}])\nonumber\\
   &=-\frac{1}{\theta_2 T_f B}\ln(\mathbb{E}[(1+ \frac{\rho P_2 |h_2|^2}{1+ \rho P_1 |h_1|^2})^{-\frac{\theta_2 T_f B}{\ln2}}])\nonumber\\
   &=-\frac{1}{\theta_2 T_f B}(\mathbb{E}[-\frac{\theta_2 T_f B}{\ln2} \ln(1+ \frac{\rho P_2 |h_2|^2}{1+ \rho P_1 |h_1|^2})])\nonumber.
\end{align}
When $\theta_2\rightarrow0$, we get an indeterminate form $\frac{0}{0}$, by applying the L'Hopital's rule one can get: 
\begin{equation}
   E_c^2 =-\frac{1}{T_f B}(\mathbb{E}[-\frac{ T_f B}{\ln2} \ln(1+ \frac{\rho P_2 |h_2|^2}{1+ \rho P_1 |h_1|^2})])\\=\mathbb{E}[\frac{1}{\ln2} \ln(1+ \frac{\rho P_2 |h_2|^2}{1+ \rho P_1 |h_1|^2})]=\mathbb{E}[\log_2(1+ \frac{\rho P_2 |h_2|^2}{1+ \rho P_1 |h_1|^2})]\nonumber
\end{equation} 
So,
\begin{equation}
   \lim\limits_{\theta_2\rightarrow0}E_c^2 =\mathbf{E}[\log_2(1+ \frac{\rho P_2 |h_2|^2}{1+ \rho P_1 |h_1|^2})]\nonumber
\end{equation} 
which is equals to $\mathbf{E}[R_2]$, the ergodic capacity.
Then,
$\lim\limits_{\theta_2\rightarrow0}E_c^2=\mathbf{E}[R_2]$, where $R_2$ is the achievable Rate. \newline
Proceeding in the same way, one can find:

\begin{equation}
\lim\limits_{\theta_1\rightarrow0}E_c^1=\mathbb{E}[\log_2(1+\rho P_1 |h_1|^2)]=\mathbb{E}[R_1]\nonumber
\end{equation} 

\begin{equation}
\lim\limits_{\theta_1\rightarrow0}\widetilde{E}_c^{1}=\mathbb{E}[\frac{1}{2}\log_2(1+\rho |h_1|^2)]=\mathbb{E}[\widetilde{R}_{1}]\nonumber
\end{equation} 

\begin{equation}
\lim\limits_{\theta_2\rightarrow0}\widetilde{E}_c^{2}=\mathbb{E}[\frac{1}{2}\log_2(1+\rho |h_2|^2)]=\mathbb{E}[\widetilde{R}_{2}]\nonumber
\end{equation} 

and,
\begin{equation}
\lim\limits_{\theta_1\rightarrow0}(E_c^1-\widetilde{E}_c^{1})=\mathbb{E}[R_1]-\mathbb{E}[\widetilde{R}_{1}]\nonumber
\end{equation}
\begin{equation}
\lim\limits_{\theta_2\rightarrow0}(E_c^2-\widetilde{E}_c^{2})=\mathbb{E}[R_2]-\mathbb{E}[\widetilde{R}_{2}]\nonumber
\end{equation}

To look further the impact of the transmit SNR $\rho$ on the EC considering delay-Unconstrained user:
 
\begin{equation}
   \lim\limits_{\substack{\theta_1\rightarrow0\\\rho\rightarrow\infty}}E_c^1 =\lim\limits_{\substack{\rho\rightarrow\infty}}\mathbb{E}[\log_2(1+\rho P_1 |h_1|^2)]=\infty\nonumber
\end{equation} 
We also have that:
\begin{equation}
   \lim\limits_{\substack{\theta_2\rightarrow0\\\rho\rightarrow\infty}}E_c^2 =\lim\limits_{\substack{\rho\rightarrow\infty}}\mathbb{E}[\log_2(1+ \frac{\rho P_2 |h_2|^2}{1+ \rho P_1 |h_1|^2})]\\=\mathbb{E}[\log_2(1+ \frac{P_2 |h_2|^2}{P_1 |h_1|^2})]\nonumber
\end{equation}
By doing the same thing we have for EC OMA:

$
   \lim\limits_{\substack{\theta_1\rightarrow0\\\rho\rightarrow\infty}}\widetilde{E}_c^{1}=\lim\limits_{\substack{\rho\rightarrow\infty}}\mathbb{E}[\frac{1}{2}\log_2(1+\rho |h_1|^2)]=\infty\nonumber
$

$
   \lim\limits_{\substack{\theta_2\rightarrow0\\\rho\rightarrow\infty}}\widetilde{E}_c^{2}=\lim\limits_{\substack{\rho\rightarrow\infty}}\mathbb{E}[\frac{1}{2}\log_2(1+\rho |h_2|^2)]=\infty\nonumber
$

And then,
$
   \lim\limits_{\substack{\theta_2\rightarrow0\\\rho\rightarrow\infty}}(E_c^2-\widetilde{E}_c^{2}) =-\infty\nonumber
$ \newline

\begin{align}
   \lim\limits_{\substack{\theta_1\rightarrow0\\\rho\rightarrow\infty}}(E_c^1-\widetilde{E}_c^{1}) &= \lim\limits_{\substack{\rho\rightarrow\infty}}(\mathbb{E}[\log_2(1+\rho P_1 |h_1|^2)]-\mathbb{E}[\frac{1}{2}\log_2(1+\rho |h_1|^2)])
   =\lim\limits_{\substack{\rho\rightarrow\infty}}(\mathbb{E}[\log2\Big(\frac{1+\rho P_1 |h_1|^2}{(1+\rho |h_1|^2)^{\frac{1}{2}}}\Big)])\nonumber\\
   &=\lim\limits_{\substack{\rho\rightarrow\infty}}(\mathbb{E}[\log2\Big(\rho^{\frac{1}{2}}\frac{(\frac{1}{\rho}+ P_1 |h_1|^2)}{(\frac{1}{\rho}+ |h_1|^2)^{\frac{1}{2}}}\Big)])= \lim\limits_{\substack{\rho\rightarrow\infty}}(\mathbf{E}[\log2\Big(\sqrt{\rho}P_1 \sqrt{|h_1|^2}\Big)])=\infty\nonumber
\end{align}

\underline{\textbf{Appendix H:}}

Using Lemma 1, when $\rho\rightarrow0$, we can show that $E_c^{1,i}-\widetilde{E}_c^{1,i}\rightarrow0$ and $E_c^{2,i}-\widetilde{E}_c^{2,i}\rightarrow0$. These terms are similar with the two-user system. Then $W_N-W_O\rightarrow0$, since 
$W_N-W_O=\sum_{i=1}^{\frac{M}{2}}(E_c^{1,i}+E_c^{2,i}-\widetilde{E}_c^{1,i}-\widetilde{E}_c^{2,i})$, \newline

And $ \lim\limits_{\rho\rightarrow0}(W_N-W_O)=0$.

On the other hand, when $\rho\rightarrow\infty$,
\begin{multline}
    W_N-W_O=\sum_{i=1}^{\frac{M}{2}}\Big(\frac{1}{\beta_{1,i}}\log_2(\mathbb{E}[(1+ \rho P_{1,i}|h_{1,i}|^2)^{\frac{2 \beta_{1,i}}{M}}])- \frac{1}{\beta_{1,i}}\log_2(\mathbb{E}[(1+ \rho |h_{1,i}|^2)^{\frac{\beta_{1,i}}{M}}])\\+\frac{1}{\beta_{2,i}}\log_2(\mathbb{E}[(1+ \frac{\rho P_{2,i}|h_{2,i}|^2}{1+ \rho P_{1,i}|h_{1,i}|^2})^{\frac{2 \beta_{2,i}}{M}}])- \frac{1}{\beta_{2,i}}\log_2(\mathbb{E}[(1+\rho |h_{2,i}|^2)^{\frac{\beta_{2,i}}{M}}])\Big)\\
    =\sum_{i=1}^{\frac{M}{2}}\Big(\frac{1}{\beta_{1,i}}\log_2(\frac{\mathbb{E}[(1+ \rho P_{1,i}|h_{1,i}|^2)^{\frac{2 \beta_{1,i}}{M}}]}{\mathbb{E}[(1+ \rho |h_{1,i}|^2)^{\frac{\beta_{1,i}}{M}}]})+\frac{1}{\beta_{2,i}}\log_2(\frac{\mathbb{E}[(1+ \frac{\rho P_{2,i}|h_{2,i}|^2}{1+ \rho P_{1,i}|h_{1,i}|^2})^{\frac{2 \beta_{2,i}}{M}}}{\mathbb{E}[(1+\rho |h_{2,i}|^2)^{\frac{\beta_{2,i}}{M}}]})\\=\sum_{i=1}^{\frac{M}{2}}\Big(\frac{1}{\beta_{1,i}}\log_2(\rho^{\frac{\beta_{1,i}}{M}}\frac{\mathbb{E}[(\frac{1}{\rho}+ P_{1,i}|h_{1,i}|^2)^{\frac{2 \beta_{1,i}}{M}}]}{\mathbb{E}[(\frac{1}{\rho}+  |h_{1,i}|^2)^{\frac{\beta_{1,i}}{M}}]})+\frac{1}{\beta_{2,i}}\log_2(\rho^{-\frac{\beta_{2,i}}{M}}\frac{\mathbb{E}[(1+ \frac{P_{2,i}|h_{2,i}|^2}{\frac{1}{\rho}+  P_{1,i}|h_{1,i}|^2})^{\frac{2 \beta_{2,i}}{M}}}{\mathbb{E}[(\frac{1}{\rho}+ |h_{2,i}|^2)^{\frac{\beta_{2,i}}{M}}]})\Big)\nonumber
\end{multline}
Then,
\begin{align}
    W_N-W_O&=\sum_{i=1}^{\frac{M}{2}}\Big(\frac{1}{\beta_{1,i}}\log_2(\frac{\mathbb{E}[(\frac{1}{\rho}+ P_{1,i}|h_{1,i}|^2)^{\frac{2 \beta_{1,i}}{M}}]}{\mathbb{E}[(\frac{1}{\rho}+  |h_{1,i}|^2)^{\frac{\beta_{1,i}}{M}}]})+\log_2(\rho^{\frac{1}{M}})-\log_2(\rho^{\frac{1}{M}})\nonumber\\
    &+\frac{1}{\beta_{2,i}}\log_2(\frac{\mathbb{E}[(1+ \frac{P_{2,i}|h_{2,i}|^2}{\frac{1}{\rho}+  P_{1,i}|h_{1,i}|^2})^{\frac{2 \beta_{2,i}}{M}}]}{\mathbb{E}[(\frac{1}{\rho}+ |h_{2,i}|^2)^{\frac{\beta_{2,i}}{M}}]})\Big)\nonumber\\
    &=\sum_{i=1}^{\frac{M}{2}}\Big(\frac{1}{\beta_{1,i}}\log_2(\frac{\mathbb{E}[(\frac{1}{\rho}+ P_{1,i}|h_{1,i}|^2)^{\frac{2 \beta_{1,i}}{M}}]}{\mathbb{E}[(\frac{1}{\rho}+  |h_{1,i}|^2)^{\frac{\beta_{1,i}}{M}}]})+\frac{1}{\beta_{2,i}}\log_2(\frac{\mathbb{E}[(1+ \frac{P_{2,i}|h_{2,i}|^2}{\frac{1}{\rho}+  P_{1,i}|h_{1,i}|^2})^{\frac{2 \beta_{2,i}}{M}}]}{\mathbb{E}[(\frac{1}{\rho}+ |h_{2,i}|^2)^{\frac{\beta_{2,i}}{M}}]})\Big)\nonumber
\end{align}

\begin{align}
  \lim\limits_{\rho\rightarrow\infty}(W_N-W_O)&=\sum_{i=1}^{\frac{M}{2}}\Big(\frac{1}{\beta_{1,i}}\log_2(\frac{\mathbb{E}[( P_{1,i}|h_{1,i}|^2)^{\frac{2 \beta_{1,i}}{M}}]}{\mathbb{E}[(  |h_{1,i}|^2)^{\frac{\beta_{1,i}}{M}}]})+\frac{1}{\beta_{2,i}}\log_2(\frac{\mathbb{E}[(1+ \frac{P_{2,i}|h_{2,i}|^2}{ P_{1,i}|h_{1,i}|^2})^{\frac{2 \beta_{2,i}}{M}}]}{\mathbb{E}[( |h_{2,i}|^2)^{\frac{\beta_{2,i}}{M}}]})\Big)\nonumber\\
  &=\sum_{i=1}^{\frac{M}{2}}\Big(\frac{1}{\beta_{1,i}}\log_2(P_{1,i}^{\frac{2 \beta_{1,i}}{M}} \mathbb{E}[(  |h_{1,i}|^2)^{\frac{\beta_{1,i}}{M}}])+\frac{1}{\beta_{2,i}}\log_2(\frac{\mathbb{E}[(1+ \frac{P_{2,i}|h_{2,i}|^2}{ P_{1,i}|h_{1,i}|^2})^{\frac{2 \beta_{2,i}}{M}}]}{\mathbb{E}[( |h_{2,i}|^2)^{\frac{\beta_{2,i}}{M}}]})\Big)\nonumber
\end{align}
Which is a constant with respect of $\rho$.\newline

Further, to analyze
$\lim\limits_{\rho\rightarrow 0}(\frac{\partial (W_N-W_O)}{\partial\rho})$ and $\lim\limits_{\rho\rightarrow\infty}(\frac{\partial (W_N-W_O)}{\partial\rho})$, we analyze $\frac{\partial W_N}{\partial\rho}$ and $\frac{\partial W_O}{\partial\rho}$ \newline

\begin{align}
   \frac{\partial W_N}{\partial\rho}&=\sum_{i=1}^{\frac{M}{2}}(\frac{\partial E_c^{1,i}}{\partial\rho}+\frac{\partial E_c^{2,i}}{\partial\rho})\nonumber\\
   &=\sum_{i=1}^{\frac{M}{2}}(\frac{1}{\beta_{1,i} \ln2}\frac{\Big(\mathbb{E}[(1+\rho P_{1,i} |h_{1,i}|^2)^{\frac{2 \beta_{1,i}}{M}}] \Big)'}{\mathbb{E}[(1+\rho P_{1,i} |h_{1,i}|^2)^{\frac{2 \beta_{1,i}}{M}}]}+\frac{1}{\beta_{2,i} \ln2}\frac{\Big(\mathbb{E}[(1+ \frac{\rho P_{2,i} |h_{2,i}|^2}{1+  \rho P_{1,i} |h_{1,i}|^2})^{\frac{2 \beta_{2,i}}{M}}]\Big)'}{\mathbb{E}[(1+ \frac{\rho P_{2,i} |h_{2,i}|^2}{1+  \rho P_{1,i} |h_{1,i}|^2})^{\frac{2 \beta_{2,i}}{M}}]})\nonumber\\
  & =\sum_{i=1}^{\frac{M}{2}}(\frac{2 P_{1,i}}{M \ln2} \frac{\mathbb{E}[|h_{1,i}|^2 (1+\rho P_{1,i} |h_{1,i}|^2)^{\frac{2 \beta_{1,i}}{M}-1}]}{\mathbb{E}[(1+\rho P_{1,i} |h_{1,i}|^2)^{\frac{2 \beta_{1,i}}{M}}]}+\frac{2 P_{2,i}}{M\ln2}\frac{\mathbb{E}[\frac{ |h_{2,i}|^2}{(1+\rho P_{1,i} |h_{1,i}|^2)^2}(1+ \frac{\rho P_{2,i} |h_{2,i}|^2}{1+  \rho P_{1,i} |h_{1,i}|^2})^{\frac{2 \beta_{2,i}}{M}-1}]}{\mathbb{E}[(1+ \frac{\rho P_{2,i} |h_{2,i}|^2}{1+  \rho P_{1,i} |h_{1,i}|^2})^{\frac{2 \beta_{2,i}}{M}}]})\nonumber
\end{align}
where $(\cdot)'$ denotes the first derivative with respect of $\rho$. 
Then,
$$\lim\limits_{\rho\rightarrow 0}(\frac{\partial W_N}{\partial\rho})=\sum_{i=1}^{\frac{M}{2}}\frac{2 P_{1,i}}{M \ln2}\mathbb{E}[|h_{1,i}|^2]+\frac{2 P_{2,i}}{M\ln2}\mathbb{E}[|h_{2,i}|^2 $$
\begin{equation}
\lim\limits_{\rho\rightarrow\infty}(\frac{\partial W_N}{\partial\rho})=\lim\limits_{\rho\rightarrow\infty}\Big(\sum_{i=1}^{\frac{M}{2}}(\frac{2}{M \ln2 \rho} +\frac{2 P_{2,i}}{M\ln2 \rho^2}\frac{\mathbb{E}[\frac{ |h_{2,i}|^2}{(P_{1,i} |h_{1,i}|^2)^2}(1+ \frac{P_{2,i} |h_{2,i}|^2}{P_{1,i} |h_{1,i}|^2})^{\frac{2 \beta_{2,i}}{M}-1}]}{\mathbb{E}[(1+ \frac{P_{2,i} |h_{2,i}|^2}{P_{1,i} |h_{1,i}|^2})^{\frac{2 \beta_{2,i}}{M}}]})\Big)=0\nonumber
\end{equation}
In the same way,
\begin{align}
   \frac{\partial W_O}{\partial\rho}&=\sum_{i=1}^{\frac{M}{2}}(\frac{\partial \widetilde{E}_c^{1}}{\partial\rho}+\frac{\partial \widetilde{E}_c^{2}}{\partial\rho})\nonumber\\
  & =\sum_{i=1}^{\frac{M}{2}}(\frac{1}{\beta_{1,i} \ln2}\frac{\Big(\mathbb{E}[(1+\rho |h_{1,i}|^2)^{\frac{ \beta_{1,i}}{M}}] \Big)'}{\mathbb{E}[(1+\rho |h_{1,i}|^2)^{\frac{ \beta_{1,i}}{M}}]}+\frac{1}{\beta_{2,i} \ln2}\frac{\Big(\mathbb{E}[(1+\rho |h_{2,i}|^2)^{\frac{ \beta_{2,i}}{M}}] \Big)'}{\mathbb{E}[(1+\rho |h_{2,i}|^2)^{\frac{ \beta_{2,i}}{M}}]})\nonumber\\
   &=\sum_{i=1}^{\frac{M}{2}}(\frac{1}{M \ln2}\frac{\mathbb{E}[|h_{1,i}|^2 (1+\rho |h_{1,i}|^2)^{\frac{ \beta_{1,i}}{M}-1}]}{\mathbb{E}[(1+\rho |h_{1,i}|^2)^{\frac{ \beta_{1,i}}{M}}]}+\frac{1}{M \ln2}\frac{\mathbb{E}[|h_{2,i}|^2 (1+\rho |h_{2,i}|^2)^{\frac{\beta_{2,i}}{M}-1}]}{\mathbb{E}[(1+\rho |h_{2,i}|^2)^{\frac{ \beta_{2,i}}{M}}]}).
\end{align}
Then we have,
$$\lim\limits_{\rho\rightarrow 0}(\frac{\partial M_0}{\partial\rho})=\sum_{i=1}^{\frac{M}{2}}\frac{1}{M \ln2}\mathbb{E}[|h_{1,i}|^2]+\frac{1}{M\ln2}\mathbb{E}[|h_{2,i}|^2 $$
and 
$$\lim\limits_{\rho\rightarrow \infty}(\frac{\partial M_0}{\partial\rho})=\lim\limits_{\rho\rightarrow \infty}\Big(\sum_{i=1}^{\frac{M}{2}}\frac{1}{\rho M \ln2}+\frac{1}{\rho M\ln2}\Big)=0$$.

So, 
\begin{align}
\lim\limits_{\rho\rightarrow0}(\frac{\partial (W_N-W_O)}{\partial\rho})&=\sum_{i=1}^{\frac{M}{2}}\Big(\frac{2 P_{1,i}}{M \ln2}\mathbb{E}[|h_{1,i}|^2]+\frac{2 P_{2,i}}{M\ln2}\mathbb{E}[|h_{2,i}|^2-\frac{1}{M \ln2}\mathbb{E}[|h_{1,i}|^2]-\frac{1}{M\ln2}\mathbb{E}[|h_{2,i}|^2\Big)\nonumber\\
&=\sum_{i=1}^{\frac{M}{2}}\Big(\frac{2 P_{1,i}-1}{M \ln2}\mathbb{E}[|h_{1,i}|^2]+\frac{2 P_{2,i}-1}{M\ln2}\mathbb{E}[|h_{2,i}|^2\Big)\nonumber\\
&=\sum_{i=1}^{\frac{M}{2}}\Big(\frac{2 P_{1,i}-1}{M \ln2}(\mathbb{E}[|h_{1,i}|^2]-\mathbb{E}[|h_{2,i}|^2)\Big)\geq0 . \nonumber
\end{align}
because of the constrain $P_1\leq\frac{1}{2}$, so $2 P_{1}\leq1$.

We get easily also, 
$$
\lim\limits_{\rho\rightarrow\infty}(\frac{\partial (W_N-W_O)}{\partial\rho})=0\nonumber$$.

\newpage
\bibliographystyle{plain}
\bibliography{biblio}

\end{document}